\documentclass{ws-ijmpd}
\usepackage[super,compress]{cite}
\begin{document}

\markboth{}
{Lightlike singular hypersurfaces in quadratic gravity}

%
\catchline{}{}{}{}{}
%

\title{Lightlike singular hypersurfaces in quadratic gravity }

\author{Victor A. Berezin}

\address{Institute for Nuclear Research of the Russian Academy of Sciences, Moscow, 117312 Russia\\
berezin@inr.ac.ru}

\author{Inna D. Ivanova}

\address{Institute for Nuclear Research of the Russian Academy of Sciences, Moscow, 117312 Russia\\
pc\textunderscore mouse@mail.ru}

\maketitle


\begin{abstract}
Using the principle of least action, the motion equations for a singular hypersurface of arbitrary type in quadratic gravity are derived. Equations containing the “external pressure” and the “external flow” components of the surface energy-momentum tensor together with the Lichnerowicz conditions serve to find the hypersurface itself, while the remaining ones define arbitrary functions that arise due to the implicit presence of the delta function derivative. It turns out that neither double layers nor thin shells exist for the quadratic Gauss-Bonnet term. It is shown that there is no “external pressure” for null singular hypersurfaces. The Lichnerowicz conditions imply the continuity of the scalar curvature in the case of spherically symmetric null singular hypersurfaces. These hypersurfaces must be thin shells if the Lichnerowicz conditions are necessary. It is shown that for this particular case the Lichnerowicz conditions can be completely removed therefore a spherically symmetric null double layer exists. Spherically symmetric null singular hypersurfaces in conformal gravity are explored as application.
\end{abstract}


\ccode{PACS numbers: \underline{02.40.Ky},  \underline{04.20.Fy}, \underline{04.50.Kd}.}


\section{Introduction}

\par It is well known that any field theory can be described via variational principle. In general relativity the metric tensor act as a dynamic variable. The action used in this variational formulation is the Einstein-Hilbert action, which is second order in terms of the metric derivatives. At the same time, this action is the simplest version that can reproduce the Einstein equations. In theory, more complex actions with higher order derivatives are possible. These theories, however, are frequently dismissed because they obey Ostrogradsky's instability theorem  \cite{Ostrogradsky}, which classifies all nondegenerate theories with higher derivatives as Lyapunov unstable. The Hamiltonian of these theories is unbounded from below. In spite of this, however, there has been a recent flurry of interest in such theories for various reasons.
\par Unlike general relativity higher derivative theories are renormalisable \cite{Hooft,Weinberg,Deser,Stelle}. Thus they might play an important role in quantizing gravity.

\par Several groups of researchers have discovered independently \cite{Utiyama,Zeldovich70,Grib,Pitaevsky,Zeldovich72,Parker,Hu,Fulling,Fulling74,Lukash,Zeldovich77} that the quantum corrections that emerge in the one-loop approximation during renormalization in any relativistic quantum field theory generate terms that are missing in in the Einstein-Hilbert action. In particular, local terms quadratic in the Riemann tensor and its contractions with coefficients of dimension of non-negative powers of energy appear. The theory that include those quadratic terms in the gravitational action is commonly known as quadratic gravity.

\par A.A. Starobinsky \cite{Starobinsky} exploited  these inevitable corrections to the gravitational Lagrangian and obtained a singularity-free cosmological solution of the de Sitter type. As a result, the first inflation model was created. 
\par For the reasons listed above we have chosen quadratic gravity as an object of study.

\par A hypersurface is called singular if the curvature tensor has singularities on it namely terms proportional to $\theta$-functions or $\delta$-functions. These hypersurfaces usually serve to describe the the local concentration of matter or energy on a given hypersurface in both general relativity and quadratic gravity. Moreover, the study of the singular distribution of matter makes it possible to construct an additional class of exact solutions describing various physical models by using metrics that are known solutions of the motion equations in the bulk for the corresponding theory of gravity.
\par In general relativity field equations on a singular hypersurface were obtained for the first time by W. Israel \cite{Israel66,Israel67,IsraelCruz}. An analogue of Israel's equations in quadratic gravity was discovered by J.M.M. Senovilla \cite{Senovilla2013,Senovilla2014,Senovilla2015,SenovillaReina,Senovilla2017,Senovilla2018}. Their main difference from the case of general relativity is in the presence of the derivative of the delta function. Thus, in quadratic gravity, in addition to thin shells, a new type of singular hypersurfaces called double layer appears.
\par The paper Ref.~\refcite{Berezin2020} demonstrated that the field equations on a singular hypersurface in quadratic gravity can be obtained using the principle of least action. Such a method has a noticeable advantage: the derivative of the delta-function does not appear explicitly at all, and the delta-function itself appears only virtual. 
\par This article generalizes the results obtained in the Ref.~\refcite{Berezin2020} to the hypersurfaces of arbitrary type but the main focus is on the null hypersurfaces. A special case of spherically symmetric lightlike shells and double layers along with the possibility of weakening the Lichnerowicz conditions for this type of hypersurfaces have been investigated. In addition, null spherically symmetric singular hypersurfaces separating two spherically symmetric solutions of conformal gravity presented in the paper Ref.~\refcite{Berezin2016} are considered here.

\section{Preliminaries}

\par The action of quadratic gravity in the general case can be represented as:
\begin{multline} \label{1}
    S_{q} =-\frac{1}{16 \pi} \int_{\Omega} \sqrt{-g} L_{q}\, d^{4}x, \\ L_{q}=\alpha_1 R_{abcd} R^{abcd}+\alpha_2 R_{ab}R^{ab}+\alpha_3 R^2+\alpha_4 R+\alpha_5\Lambda .
\end{multline}
Here $g$ is the determinant of the metric whose components act as dynamic variables since we are working in the framework of Riemannian geometry, $\alpha_{i}$ are arbitrary constants, $R_{\; bcd}^{a}$ is Riemann curvature tensor:
$$
  R_{\; bcd}^{a}=\partial _{c}\Gamma _{bd}^{a}-\partial _{d}\Gamma _{bc}^{a}+\Gamma_{ce}^{a}\Gamma_{bd}^{e}-\Gamma_{de}^{a}\Gamma_{bc}^{e} \, .
$$
\par The Ricci tensor and scalar curvature are defined as follows: $R_{ab}=R_{\; acb}^{c}, \; R=R^{a}_{a}$ . Signature $(+,-,-,-)$ and geometric units in which $c=G=1$ are used in this article.
\par For certain problems, it is also convenient to represent the Lagrangian of quadratic gravity in the following form: 
\begin{multline} \label{2a}
    L_{q}=\left ( 2\alpha_{1}+\frac{1}{2} \alpha_{2}\right )C^{2}-\left ( \alpha_{1}+\frac{1}{2} \alpha_{2}\right )GB+\frac{1}{3}\left (3\, \alpha _{3}+ \alpha_{1}+ \alpha_{2}\right )R^{2}+\\+\alpha_4 R+\alpha_5\Lambda ,\quad C^2=C_{abcd} \, C^{abcd}=R_{abcd} \, R^{abcd}-2 R_{ab} \, R^{ab}+\frac{1}{3} R^2 ,\\
    GB=R_{abcd}\, R^{abcd}-4 R_{ab} \, R^{ab}+ R^2,
\end{multline}
where $ C_{abcd} $ is the Weyl tensor, which is defined as:
\begin{multline} \label{2d}
   C_{a b c d}=R_{a b c d }+\frac{1}{2}\left ( R_{a d}\, g_{b c }+R_{b c} \, g_{a d}-R_{a c}\, g_{b d}-R_{b d}\, g_{a c}\right )+\\+\frac{1}{6}R\left ( g_{a c}\, g_{b d}-g_{a d}\, g_{b c} \right ) .
\end{multline}
\par Let us consider a four-dimensional spacetime $\Omega$ divided  by the singular hypersurface $\Sigma_0$ into two regions $\Omega^{+}$ and $\Omega^{-}$ with different geometries. As stated before a singular hypersurface is a hypersurface where the Riemann curvature tensor has a singular component. 
\par Let the hypersurface equation in some coordinates continuous in the neighborhood of $\Sigma_0$ have the form: $n(x)=0$, then, without loss of generality, we can assume that:
$$
(N_{a}\, N^{a})|_{\Sigma_0}=(\partial_{a}n \, \partial^{a}n)|_{\Sigma_0}=\varepsilon .
$$
Here $N^{a}$ is the normal to the hypersurface, $\varepsilon=-1$ for a timelike hypersurface, $1$ for a spacelike one and zero for a null one.
\par Let the domain $\Omega^{-}$ correspond to negative values of the function $n(x)$ and $\Omega^{+}$ to positive values, then the outward normal to $\Sigma_0$ is defined as:
$$
N_{a}=\epsilon \partial_a n \, ,
$$
where $\epsilon=\varepsilon$ for timelike and spacelike hypersurfaces and $\epsilon=1$ for lightlike ones. Such a definition will be necessary for the application of the Stokes' theorem and is dictated by the condition: $N^{a}\partial_{a}n \geq 0 $ .
\par Taking into account the accepted notation, the metric in the entire space-time $\Omega$ can be formally expressed as a sum:
\begin{equation}\label{7}
 g_{a b}=g^{+}_{a b} \; \theta(n(x))+g^{-}_{a b} \; \theta(-n(x))=g_{a b}(\pm) \, .
\end{equation}
\par Differentiating expressions  (\ref{7}) gives rise to a term with a delta function:
$$
 \partial_c g_{a b}=\partial_c g_{a b}(\pm)+\partial _{c}n(x)\, \delta(n(x))\,  [g_{ab}] .
$$
Here and below, $\partial_c$ denotes the partial derivative, the brackets $[g_{ab}]=g^{+}_{ab}-g^{-}_{ab}$ denote the jump of the corresponding value on the surface $\Sigma_0$.
\par If $[g_{ab}]\neq 0$ then the $\delta$-function occurs in Christoffel symbols. Here we will work with the conventional theory of distributions, where the products $\delta^2(n(x))$ and $\delta(n(x))\,\theta(n(x))$ are undefined. Therefore, both in general relativity and in quadratic gravity, the continuity of the metric on $\Sigma_0$ is essential in order to avoid the appearance of $\delta^2$ in the Riemann curvature tensor and in the Lagrangian .
\par As it was shown , in particular, in the paper Ref.~\refcite{Berezin87} it is always possible to make the $g_{ab}$ tensor continuous on $\Sigma_0$ via coordinate transformations in the $\Omega^{\pm}$ domains. Thus for singular hypersurfaces in both general relativity and quadratic gravity one should either work in a similar special coordinate system or explicitly impose the restrictions $[g_{ab}]=0$.
\par If the metric is continuous on $\Sigma_0$, then from (\ref{7}) we get the following expression for the Levi-Civita connection components:
\begin{equation}\label{8}
\Gamma _{bc}^{a}=\Gamma _{bc}^{a}(\pm),
\end{equation}
from where, in turn, the formula for the Riemann tensor is derived:
\begin{equation}\label{9}
  R_{\; bcd}^{a}=R_{\: bcd}^{a}(\pm )+\left \{ \partial _{c}n(x)\, [\Gamma_{bd}^{a}]-\partial _{d}n(x) \, [\Gamma _{bc}^{a}] \right \}\delta (n(x)) \, . 
\end{equation}
\par In case of general relativity the singular part of the curvature tensor corresponds to the surface part of the energy-momentum tensor \cite{Israel66,Israel67,IsraelCruz} therefore the expression  (\ref{9}) is acceptable. 
\par In quadratic gravity it is also necessary to require the continuity of the first derivatives of the metric and the Christoffel symbols as a consequence in order to avoid the appearance of undefined functions in the Lagrangian.The corresponding restrictions are called Lichnerowicz conditions \cite{Lake}:
\begin{equation} \label{10}
  [\Gamma _{bc}^{a}]=0 \, .
\end{equation}
\par For all physical models that we consider in this paper, the energy-momentum tensor of matter fields has the following structure:
\begin{equation}\label{e}
 T^{a b}=S^{a b} \; \delta(n(x))+T^{+ a b}  \; \theta(n(x))+T^{- a b}  \;  \theta(-n(x))=S^{a b} \; \delta(n(x))+T^{a b}(\pm),
\end{equation}
where $S^{ab}$ is the surface energy-momentum tensor, $T^{\pm ab}$ are the values of the energy-momentum tensor components in the domains $\Omega^{\pm}$ respectively. Here we assume that $T^{ab}$ does not contain other singular terms, in particular, derivatives of the $\delta$-function.

\section{Thin shells in General Relativity}
\par Before proceeding to the study of singular hypersurfaces in quadratic gravity, we will demonstrate how the Israel equations can be derived using the principle of least action.
\par The gravitational part of the action in this case is:
$$
S_{GR} =-\frac{1}{16 \pi} \int_{\Omega} \sqrt{|g|}L_{GR}\,  d^{4}x=-\frac{1}{16 \pi} \int_{\Omega} \sqrt{|g|}\left ( \alpha_{4}\, R +\alpha_{5}\, \Lambda \right )\,  d^{4}x \, .
$$
\par Using the expression  (\ref{9}) for the Riemann tensor, we obtain the scalar curvature for a manifold with a singular hypersurface:
$$
 R=R(\pm)+g^{bd} \, E_{bd} \,\delta (n(x)) \, .
$$
Here, for convenience, an additional tensor is introduced:  $E_{bd}=\partial _{a}n\, [\Gamma_{bd}^{a}]-\partial _{d}n \, [\Gamma _{ba}^{a}]$.
\par Next, we move on to varying the action for gravity with respect to the inverse metric:
$$
\delta\,  S_{GR} =-\frac{1}{16 \pi} \int_{\Omega} \sqrt{|g|}\left ( \delta L_{GR}-\frac{1}{2}g_{ac}\, L_{GR}\, \delta g^{ac} \right )\,  d^{4}x .
$$
Let us write out separately the variation of the scalar curvature:
\begin{multline} \nonumber
\delta R=\delta g^{bd}\left ( R_{bd}(\pm)+ E_{bd}\, \delta (n(x))\,  \right )+\\+g^{bd}\left ( \delta R_{bd} (\pm)+\left \{ \partial _{a}n\, [\delta \Gamma_{bd}^{a}]-\partial _{d}n \, [\delta \Gamma _{ba}^{a}] \right \}\, \delta (n(x))\, \right ) \, .  
\end{multline}
\par The hypersurface equation $n(x)=0$ is assumed to be unknown, but given during the variation process, i.e. the function $n(x)$ itself does not vary. Using the corollary of the Palatini formula \cite{Palatini} we get:
$$
g^{bd} \delta R_{bd}(\pm)=g^{bd}\left(\triangledown _{a}\delta \Gamma_{bd}^{a}(\pm) -\triangledown _{d}\delta \Gamma_{ab}^{a}(\pm)\right) \, . 
$$
If we then write this term from $\delta R$ as part of the original integral:
\begin{multline} \nonumber
\int_{\Omega} \sqrt{|g|} g^{bd} \delta R_{bd}(\pm)\,  d^{4}x=\int_{\Omega^{-}} \sqrt{|g|} g^{bd} \left(\triangledown _{a}\delta \Gamma_{bd}^{a} -\triangledown _{d}\delta \Gamma_{ab}^{a} \right)\,  d^{4}x+\\+\int_{\Omega^{+}} \sqrt{|g|} g^{bd} \left(\triangledown _{a}\delta \Gamma_{bd}^{a} -\triangledown _{d}\delta \Gamma_{ab}^{a}\right)\,  d^{4}x\, ,   
\end{multline}
and apply the Stokes' theorem for each of the integrals, then it turns out that this term cancels out with $\int_{\Omega} \sqrt{|g|} g^{bd} \left \{ \partial _{a}n\, [\delta \Gamma_{bd}^{a}]-\partial _{d}n \, [\delta \Gamma _{ba}^{a}] \right \}\, \delta (n(x))\,  d^{4}x$. Taking into account all of the above, we get:
$$
\int_{\Omega} \sqrt{|g|}\delta R\,  d^{4}x=\int_{\Omega} \sqrt{|g|}\, \delta g^{bd}\left (\, R_{bd}(\pm)+ E_{bd}\, \delta (n(x))\,  \right )\,  d^{4}x\, .
$$
Substituting the resulting expression into a variation of the action of gravity, we obtain:
\begin{multline}
\delta\,  S_{GR} =-\frac{1}{16 \pi} \int_{\Omega} \sqrt{|g|}\, \delta g^{ab}\left ( \alpha _{4}\, R_{ab}(\pm ) -\frac{1}{2}\alpha _{4}\, g_{ab}\, R(\pm )-\frac{1}{2}\alpha_{5}\, g_{ab}\, \Lambda \right )  d^{4}x+\\+\frac{1}{16 \pi} \int_{\Omega} \sqrt{|g|}\, \delta g_{ab} \, \alpha _{4} \left ( E^{ab} -\frac{1}{2}\, g^{ab}\, E\right )\, \delta(n(x))\,   d^{4}x\, ,
\end{multline}
where $E=g^{ab}\, E_{ab}$.
\par Similarly, for the action of matter, taking into account the structure of the energy-momentum tensor  (\ref{e}), we have:
$$
\delta S_{m} =\frac{1}{2} \int_{\Omega} \sqrt{|g|} \, T_{ab}(\pm )\, \delta g^{ab} \, d^{4}x -\frac{1}{2} \int_{\Omega } \sqrt{|g|}\, S^{ab}\, \delta g_{ab}\,  \delta (n(x))\, d^{4}x \, .
$$
\par According to the principle of least action: $\delta\, S_{GR}=-\delta S_{m}$, from this we obtain a system of equations of motion:
$$
\alpha _{4}\, R^{\pm }_{ab} -\frac{1}{2}\alpha _{4}\, g_{ab}\, R^{\pm }-\frac{1}{2}\alpha_{5}\, g_{ab}\, \Lambda=8\pi \, T^{\pm }_{ab}\, ,
$$
$$
E^{ab}-\frac{1}{2}\, g^{ab}\, E =\frac{8\pi }{\alpha _{4}} \, S^{ab}\, .
$$
\par Once the variation is done and the equations of motion are obtained, it is possible to move to specific coordinates. We choose a coordinate system $\{n, y^i\}$, where the function $n(x)$, which defines the hypersurface equation, is one of the coordinates, $y^i$ denotes all coordinates except $n$. In addition, we require that the components of the metric in these coordinates be continuous on the hypersurface. An example of such coordinates can be Gaussian normal coordinates in the case of a timelike (spacelike) hypersurface and the coordinate systems described in the fifth section for a lightlike one. 
\par A consequence of the standard relations for the transformation of the metric components when the coordinates are changed from arbitrary to $\{n, y^i\}$ is the following formula:
$$
g^{nn}|_{\Sigma_0}=(\partial_{a}n \partial^{a}n)|_{\Sigma_0}=\varepsilon .
$$
\par Let us write out the components of the tensor $E_{ab}$ and the jumps in Christoffel symbols in the coordinates $\{n, y^i\}$:
$$
E_{bd}=[\Gamma _{bd}^{n}]-\delta _{d}^{n}\,\delta _{b}^{n} \, [\Gamma _{an}^{a}],
$$
$$
[\Gamma _{bc}^{a}]=\frac{1}{2}\, g^{ad}\left (\delta_{c}^{n} \, [\partial_{n} g_{bd}] +\delta_{b}^{n} \, [\partial_{n} g_{cd}]\right )-\frac{1}{2}g^{an}[\partial _{n}g_{bc}], \quad [\Gamma _{ba}^{a}]=\frac{1}{2}g^{ad}[\partial _{n}g_{ad}]\, \delta _{b}^{n},
$$
$$
E= \left (g^{bn}\, g^{nd} - g^{nn}\, g^{bd}  \right )[\partial _{n}g_{bd}]=\left ( g^{kn}\, g^{nl} - g^{nn}\, g^{kl} \right )[\partial _{n}g_{kl}],\quad k,l\neq n.
$$
It follows from the relations presented above that $E^{an}=\frac{1}{2}\,g^{an}\,E$, therefore the components $S^{nn}$ and $S^ {ni}$ of the surface energy-momentum tensor are zero in general relativity.
\par The motion equations of a singular hypersurface in the coordinates  $\{n, y^i\}$ are:
\begin{equation} \label{eqgr}
\left (g^{ai}g^{bj}-\frac{1}{2}\, g^{ij}g^{ab}\right )[\Gamma _{ab}^{n}]+\left (\frac{1}{2}\, g^{ij} g^{nn}-g^{ni}g^{nj}\right )[\Gamma _{cn}^{c}]=\frac{8\pi }{\alpha _{4}} \, S^{ij}\, ,  \quad i,j \neq n\, ,
\end{equation}
in this form, they can be used not only for timelike (spacelike) hypersurfaces but also for lightlike hypersurfaces.
\par A special case of (\ref{eqgr}) are the Israel equations for a timelike (spacelike) thin shell. In a Gaussian normal coordinate system, where the metric in the vicinity of the hypersurface has the form: $ds^2=\varepsilon dn^2+\gamma_{ij}\, dx^{i}\,dx^{j}$, motion equations (\ref{eqgr}) come down to the following:
$$
\varepsilon \left ( [K^{ij}]-\gamma ^{ij}[K]\right )=\frac{8\pi }{\alpha _{4}}\, S^{ij}\, ,
$$
where $K_{ij}=-\frac{1}{2}\partial_{n}\,\gamma_{ij}$ is the external curvature tensor.
\par Let us also consider the case of null hypersurfaces. For this purpose we use the special coordinates $\{n,\lambda, \theta^A\}$ described in the fifth section. In these coordinates the equations of motion (\ref{eqgr}) are:
\begin{equation} \label{eqnull}
\frac{1}{2}\left (\delta _{\lambda }^{i}\,  g^{aj} +\delta _{\lambda }^{j}\,  g^{ai}\right )[\partial _{n}g_{a\lambda }]-\frac{1}{2}g^{ij}[\partial _{n}g_{\lambda \lambda }]- \frac{1}{2} \delta _{\lambda }^{i}\, \delta _{\lambda }^{j}\, g^{ab} [\partial _{n} g_{ab}]=\frac{8\pi }{\alpha _{4}} \, S^{ij}\, .   
\end{equation}
\par In addition, let's explore the spherically symmetric null thin shell. As it will be shown below, the metric in the neighborhood of such a hypersurface has the form: $ds^2=\gamma_{nn}\,dn^2+2\gamma_{n\lambda}\,dn d\lambda-r^2\,d\Omega^2$. Therefore the motion equations for a null thin shell (\ref{eqnull}) come down to the following in the spherically symmetric case:
$$
[\partial _{n}r]=-\frac{4\pi r^{2} }{\alpha _{4}} \, S^{\lambda \lambda }\, .
$$
This formula is fully consistent with the result obtained in the article Ref.~\refcite{Berezin87}.

\section{Motion equations}
\par The main purpose of this paper is to derive the field equations for a singular null hypersurface in quadratic gravity. In order to achieve it we need to isolate the surface part in the motion equations obtained by varying the action  (\ref{1}) with respect to the inverse metric in the case when the curvature tensor has a structure described by (\ref{9}) with (\ref{e}) as a source. 

\par It was shown in the previous section that for a singular hypersurface of any type in general relativity, the components of the surface energy-momentum tensor $S^{nn}$ and $S^{ni}$ are equal to zero. In the case of quadratic gravity, they are generally nonzero. It was shown for the first time in the papers of J. M. M. Senovilla \cite{Senovilla2013,Senovilla2014,Senovilla2015,SenovillaReina,Senovilla2017,Senovilla2018}, where $S^{nn}$ and $S^{ni}$ are defined as “external pressure” and “external flow” respectively. 

\par In general relativity, the field equations are of the second order in the derivatives of the metric tensor. If 
$[\Gamma _{bc}^{a}]\neq 0$ and a $\delta$-function arises in the Riemann tensor as well as in the energy-momentum tensor then the Israel equations link these jumps to $S^{ab}$. The hypersurface is a thin shell in this case. If $[\Gamma _{bc}^{a}]= 0$ there is only a jump in both $T^{ab}$ and curvature tensor then there is a gravitational shock wave accompanied by a shock wave in matter.

\par In quadratic gravity, the field equations are of the fourth order in the derivatives of the metric tensor. In the case of thin shell the curvature tensor doesn't have a jump on $\Sigma_0$ so its second derivatives can contain at most $\delta$-function, which relates to the $\delta$-function in the energy-momentum tensor. In the case of double layer discovered by J. M. M. Senovilla the curvature tensor experience a jump on the hypersurface, then its second derivatives contain a derivative of $\delta$-function. The curvature jump describing the gravitational shock wave doesn't have to be accompanied by a shock wave in the distribution of matter, thus the existence of purely gravitational shock wave is possible in quadratic gravity.

\par Due to the Lichnerowicz conditions the Riemann tensor and, as a consequence, the Ricci tensor and scalar curvature may have at most a jump on the $\Sigma_0$. Using the fact that $\theta^2(n(x))=\theta(n(x))$ and $\theta(n(x)) \, \theta(-n(x))=0$, and substituting the Riemann tensor  (\ref{9}) and its contractions into the action (\ref{1}), we get:
\begin{equation} \label{11}
    S_{q} =-\frac{1}{16 \pi} \int_{\Omega^{+}} \sqrt{|g|} L^{+}_{q}\, d^{4}x-\frac{1}{16 \pi} \int_{\Omega^{-}} \sqrt{|g|} L^{-}_{q}\, d^{4}x,
\end{equation}
where $L^{+}_{q}=\alpha_1 R^{+}_{abcd} R^{+abcd}+\alpha_2 R^{+}_{ab} R^{+ab}+\alpha_3 (R^{+})^{2}+\alpha_4 R^{+}+\alpha_5\Lambda$, and $L^{-}_{q}$ is defined in a similar way. 
\par Varying the action  (\ref{11}) with respect to the inverse metric, we obtain \cite{Deruelle2003}:
\begin{multline} \label{12}
    \delta S_{q} =-\frac{1}{16 \pi} \int_{\Omega^{+}} \sqrt{|g|} \left(H^{+}_{ab}\delta g^{ab}+\triangledown_c V^{+c}\right) \, d^{4}x-\\-\frac{1}{16 \pi} \int_{\Omega^{-}} \sqrt{|g|} \left(H^{-}_{ab}\delta g^{ab}+\triangledown_c V^{-c}\right) \, d^{4}x .
\end{multline}
The following notation is adopted here:
\begin{multline} \nonumber
   H^{+}_{ab}=2\alpha_1 R^{+}_{amlp}R_{b}^{+ mlp}-2(2\alpha_1+\alpha_2) R^{+c}_{d} R^{+d}_{\;\;bac}-4\alpha_1 R^{+c}_{a}R^{+}_{bc}+2\alpha_3 R^{+}R^{+}_{ab}+\alpha_4 R^{+}_{ab}-\\-\frac{1}{2}g_{ab}L^{+}_{q}+(\alpha_2+4\alpha_1)\square R^{+}_{ab}+\frac{1}{2}(4\alpha_3+\alpha_2)g_{ab}\square R^{+}-(2\alpha_1+\alpha_2+2\alpha_3)\triangledown_a \triangledown_b R^{+},
\end{multline}
\begin{multline} \nonumber
 V^{+c}=\left \{ (4\alpha_{1}+\alpha_{2})\triangledown^{c} R^{+bd}+\frac{1}{2}(\alpha_{2}+4\alpha_{3})g^{bd}\triangledown^{c}R^{+} \right\}\delta g_{bd}-\\-2\triangledown^{b}\left \{ (2\alpha _{1}+\alpha _{2})R^{+cd}+\alpha_{3} \, g^{cd}R^{+} \right \}\delta g_{bd} +(\alpha_{4}+2\alpha_{3}R^{+})(g^{ab}g^{cd}-g^{ac}g^{bd})\, \triangledown_{a} \delta g_{bd}+\\+\left \{\alpha_{2}(2g^{cd}R^{+ab}-g^{ac}R^{+bd}-g^{bd}R^{+ac})-4\alpha_{1}R^{+abcd}  \right \}\triangledown_{a} \delta g_{bd} \, .
\end{multline}
Vector $V^{-c}$ and tensor $H^{-}_{ab}$ are defined in a similar way. 
\par By definition variation with fixed ends is used to derive equations of motion from the principle of least action. This means that the values of dynamic variables are fixed on the boundary of the entire integration volume: $\delta g_{ab}=0$ on $\partial \Omega$. Using the Stokes' theorem, we obtain the following expression for $\delta S_{q}$:
\begin{equation} \nonumber
    \delta S_{q} =-\frac{1}{16 \pi} \int_{\Omega^{+}} \sqrt{|g|} \left(H^{+}_{ab}\delta g^{ab}\right) \, d^{4}x-\frac{1}{16 \pi} \int_{\Omega^{-}} \sqrt{|g|} \left(H^{-}_{ab}\delta g^{ab}\right) \, d^{4}x +\frac{1}{16 \pi} \int_{\Sigma_0} [V^{c}] \, dS_c .
\end{equation}
Here $dS_c$ is a directed surface element.
\par As in the previous section, we use the coordinates $\{n, y^i \}$, where the metric is continuous on $\Sigma_0$ and the function $n(x)$ is one of the coordinates. In such a coordinate system we can use the definition of $dS_c$ that is applicable to both timelike (spacelike) and null cases:
\begin{equation} \nonumber
    dS_{c}=\epsilon N_{c} \sqrt{|h|} d^3 y, \quad i \neq n,
\end{equation}
where $h=g(0,y^i)$ is the restriction of the entire spacetime $\Omega$ metric determinant to $\Sigma_0$, $\left\{ y^i \right\}$ are all coordinates except $n$, they are chosen as internal coordinates on the hypersurface in this case. The directed surface element defined in this way satisfies the relation:
$$
N^{c}dS_{c}\geq 0 .
$$
\par The variation of the matter action defined by the energy-momentum tensor  (\ref{e}) is divided into volume and surface parts in a similar way:
\begin{multline} \nonumber
    \delta S_{m} =\frac{1}{2} \int_{\Omega^{+}} \sqrt{|g|} \left(T^{+}_{ab}\delta g^{ab}\right) \, d^{4}x+\frac{1}{2} \int_{\Omega^{-}} \sqrt{|g|} \left(T^{-}_{ab}\delta g^{ab}\right)\, d^{4}x-\\ -\frac{1}{2} \int_{\Sigma_0} \sqrt{|h|} \left(S^{ab}\delta g_{ab}\right) \, d^{3}y .
\end{multline}
\par From the principle of least action we obtain the system of motion equations:
\begin{equation} \label{18}
    H^{\pm}_{ab}=8\pi T^{\pm}_{ab},
\end{equation}
\begin{equation} \label{19}
 \epsilon [V^{c}] N_{c}= 8\pi S^{ab}\delta g_{ab} .
\end{equation}
\par From  (\ref{19}) an analogue of the Israel equation for quadratic gravity can be obtained. The case of timelike and spacelike hypersurfaces is considered, for example, in Ref.~\refcite{SenovillaReina} and Ref.~\refcite{Berezin2020}. Here we focus on the null case. However, the expression  (\ref{19}) can be greatly simplified before moving on to a specific type of hypersurface.
\par Since the variation of the gravitational part of the action $\delta S_q$ contains a total divergence from the vector $V^c$, this vector is defined up to the addition of another divergence-less vector $2 C U^{c}$. It can be shown \cite{Deruelle2003} that in this case it is convenient to choose $U^c$ as follows:
\begin{equation} \nonumber
U^{c}=-\triangledown^{b} \left(R^{cd}-\frac{1}{2} g^{cd} R\right) \delta g_{bd}+\left(R^{ab} g^{cd}-R^{bc} g^{ad}\right) \triangledown_{a} \delta g_{bd} .
\end{equation}
Also we set the constant $C$ equal to $2\alpha_3$, then:
\begin{multline} \nonumber
 \widetilde{V}^{+c}=V^{+c}+4\alpha_3 U^{+c}=\left \{ (4\alpha_{1}+\alpha_{2})\triangledown ^{c} R^{+bd}+\frac{1}{2}(\alpha_{2}+4\alpha_{3})g^{bd}\triangledown^{c}R^{+} \right \}\delta g_{bd}-\\-2(2\alpha _{1}+\alpha _{2}+2\alpha_3)\triangledown^{b} R^{+cd} \delta g_{bd}+(\alpha_{4}+2\alpha_{3}R^{+})(g^{ab}g^{cd}-g^{ac}g^{bd})\triangledown_{a} \delta g_{bd}\\+\left \{\alpha_{2}(2g^{cd}R^{+ab}-g^{ac}R^{+bd}-g^{bd}R^{+ac})-4\alpha_{1}R^{+abcd}  \right \}\triangledown_{a} \delta g_{bd}+\\+4\alpha_3 (R^{+ab} g^{cd}-R^{+bc} g^{ad})\triangledown_{a} \delta g_{bd}\,  .
\end{multline}
As before, the vectors $U^{-c}$,$\widetilde{V}^{-c}$ are defined similarly to $U^{+c}$, $\widetilde{V}^{+c}$ .

\par Let us explain why adding the vector $U^c$ does not change motion equations not only in the bulk but also on the hypersurface. As noted in Ref.~\refcite{Deruelle2003}, the vector $U^c$ in fact is a total divergence:
$$
U^{c}=\triangledown _{a}\left \{ \left (R^{ab}g^{cd} -R^{cd}g^{ab} \right ) \delta g_{bd}\right \} .
$$
It follows from the formula  (\ref{22}) that only the component $U^n$ contribute to the motion equations and it can be further simplified:
\begin{multline} \nonumber
  U^{n}=\triangledown _{n}\left \{ \left (R^{nb}g^{nd} -R^{nd}g^{nb} \right ) \delta g_{bd}\right \}+\triangledown _{i}\left \{ \left (R^{ib}g^{nd} -R^{nd}g^{ib} \right ) \delta g_{bd}\right \}=\\=\triangledown _{i}\left \{ \left (R^{ib}g^{nd} -R^{nd}g^{ib} \right ) \delta g_{bd}\right \},\quad i\neq n . 
\end{multline}
\par Let us show that $\int_{\Sigma_0} \triangledown _{i}\left \{ \left (R^{ib}g^{nd} -R^{nd}g^{ib} \right ) \delta g_{bd}\right \} \, \sqrt{|h|} d^3 y$ is zero. Since the expression under the integral sign is the total divergence of some three-dimensional vector field on the $\Sigma_0$, this integral reduces to the integral over the boundary of the hypersurface due to the Stokes' theorem. The boundary of $\Sigma_0$ is a part of the entire spacetime boundary $\partial \Omega$ where $\delta g_{bd}=0$ that is why the integral is equal to zero. 
\par In the coordinates $\left\{n,y^i \right\}$: $N_{c}=\epsilon \delta^{n}_{c}$, so replacing $V^{\pm c}$ with $\widetilde{V}^{\pm c}$ in  (\ref{19}) we get:
\begin{equation} \label{22}
 [\widetilde{V}^{n}] = 8\pi S^{ab}\delta g_{ab} .
\end{equation}
\par If the Lichnerowicz conditions are satisfied, then $[\triangledown_{a} \delta g_{bd}]=0$ as a consequence so this factor can be taken out of the bracket when calculating the jump $[\widetilde{V}^{n}]$ :
\begin{multline} \label{23}
[\widetilde{V}^{n}]=\left \{ (4\alpha_{1}+\alpha_{2})[\triangledown ^{n} R^{bd}]+\frac{1}{2}(\alpha_{2}+4\alpha_{3})g^{bd}[\triangledown^{n}R] \right \}\delta g_{bd}+\\+\left \{ 2\alpha_{3} [R](g^{ab}g^{n d}-g^{a n}g^{bd})+\alpha_{2}(2g^{n d}[R^{ab}]-g^{a n}[R^{bd}]-g^{bd}[R^{a n}])\right \}\triangledown_{a} \delta g_{bd}+\\+\left \{ 4\alpha_3 ([R^{ab}] g^{n d}-[R^{b n}] g^{ad})-4\alpha_{1}[R^{ab n d}]  \right \}\triangledown_{a} \delta g_{bd}-\\-2(2\alpha _{1}+\alpha _{2}+2\alpha_3)[\triangledown^{b} R^{n d}] \, \delta g_{bd}.
\end{multline}
\par It is necessary to obtain expressions for the jumps of the Riemann tensor and its contractions in order to calculate $[\widetilde{V}^{n}]$, keeping in mind that jumps are present only in those quantities that contain derivatives of the metric with respect to the coordinate $n$, starting with the second order derivative. The jumps of the first derivatives of the metric are set to zero due to the  Lichnerowicz conditions:
\begin{equation} \label{24}
[R_{iklm}]=\frac{1}{2}\left(\delta^{n}_{k} \delta^{n}_{l}[\partial^{2}_{n n}g_{im}]+\delta^{n}_{i} \delta^{n}_{m}[\partial^{2}_{n n}g_{kl}]-\delta^{n}_{i} \delta^{n}_{l}[\partial^{2}_{n n}g_{km}]-\delta^{n}_{k} \delta^{n}_{m}[\partial^{2}_{n n}g_{il}]\right),
\end{equation}
\begin{equation}
[R^{abcd}\, ]=g^{bn}\, g^{cn}\, [R_{\, \, nn\, }^{a\, \,\, \, \,  d}\, ]+g^{an}\, g^{dn}\, [R_{\, \, nn\, }^{c\, \,\, \, \,  b}\, ]-g^{bn}\, g^{dn}\, [R_{\, \, nn\, }^{a\, \,\, \, \,  c}\, ]-g^{an}\, g^{cn}\, [R_{\, \, nn\, }^{b\, \,\, \, \,  d}\, ]
\end{equation}

\begin{multline} \label{26}
[ R^{ab}]=\frac{1}{2}\left (- g^{an} g^{bn} g^{cd}-g^{nn}g^{ac}g^{bd}+g^{an} g^{bd} g^{cn}+g^{bn} g^{ad} g^{cn} \right )[\partial_{nn}^{2} g_{cd}]=\\ =g^{bn}\, [R_{\, \, nn\, }^{n\, \,\, \, \,  a}\, ]+g^{an}\, [R_{\, \, nn\, }^{n\, \,\, \, \,  b}\, ]-g^{nn}\, [R_{\, \, nn\, }^{a\, \,\, \, \,  b}\, ]+g^{an}\, g^{bn}\, [R_{nn}],
\end{multline}
\begin{equation} \label{27}
[R]=\left (-g^{nn}g^{cd}+g^{cn} g^{dn} \right )[\partial _{nn}^{2}g_{cd}]=2 \, [R_{\, \, nn\, }^{n\, \,\, \, \,  n}\, ]+2\, g^{nn}\, [R_{nn}]\, .
\end{equation}
\par The factor in $\widetilde{V}^{\pm n}$ at $\triangledown_{a} \delta g_{bd}$ (we denote it by $A^{\pm abd}$) requires a more detailed consideration, taking into account the relations  (\ref{24}-\ref{27}) we get:
\begin{multline} \label{28}
[\, A^{abd}\, ]\, \triangledown _{a}\delta g_{bd}=\left \{  \right.-\beta_{2}\,  g^{an} g^{bd} [R_{\: nn}^{n\: \:\:  n}]+ \beta_{2}\, g^{an}(g^{bn} g^{dn}-g^{bd} g^{nn})[R_{nn}]+\\+(\beta _{1}+\beta _{2})g^{nb}g^{nd}[R_{\: nn}^{n\: \:\:  a}]+(\beta _{2}-\beta _{1})g^{an}g^{dn}[R^{\: bn}_{n\: \:\:  n}]+\\+(\beta _{2}+\beta _{1})g^{nn}g^{nd}[R_{n \: \: n}^{\: \: b \: \:  a}]-\beta _{1}\, g^{nn}g^{na}[R_{n \: \: n}^{\: \: b \: \:  d}]\left.  \right \}\triangledown _{a}\delta g_{bd}=\\=\left \{  \right.-\frac{1}{2} \, \beta_2 \, g^{an} g^{bd} [R]+\frac{1}{2}(\beta_1+\beta_2)\,\left (g^{nd}[R^{ab}]+g^{nb}[R^{ad}]  \right )-\\-\beta_1\,  g^{an}[R^{bd}] \left.  \right \}\,\triangledown _{a}\delta g_{bd} \,  , \quad \beta _{1}=\alpha_2+4\alpha_1, \;  \beta _{2}=\alpha_2+4\alpha_3\, .
\end{multline}
 \par Since in general: $\triangledown_{i}[A^{ibd}]\neq [\triangledown_{ i}A^{ibd}]$, but $\partial_{i}[A^{ibd}]= [\partial_{i}A^{ibd}]$ for any $i \neq n$, it makes sense to get rid of covariant derivatives $\triangledown _{a}\delta g_{bd}$:
 $$
 [A^{abd}]\, \triangledown _{a}\delta g_{bd}=[A^{nbd}]\, \partial _{n}\delta g_{bd}+[A^{ibd}]\, \partial _{i}\delta g_{bd}-[A^{abd}]\, \Gamma _{ab}^{c}\, \delta g_{cd}-[A^{abd}]\, \Gamma _{ad}^{c}\, \delta g_{cb} ,\quad i\neq n\, .
 $$
\par In order to perform the following operation with $[A^{abd}]$, it is necessary to return to the integration of $[A^{abd}]$ as part of $[\widetilde{V}^{n}]$ over the hypersurface $\Sigma_0$:
\begin{multline} \nonumber
 \int_{\Sigma_0} [A^{ibd}]\, \partial _{i}\delta g_{bd} \, \sqrt{|h|} \, d^3 y=\int_{\Sigma_0} \partial _{i}\left( [A^{ibd}]\,\delta g_{bd} \, \sqrt{|h|} \right) \, d^3 y-\\- \int_{\Sigma_0}\,\left ( [\partial _{i}A^{ibd}] +[A^{ibd}]\, \frac{1}{2}\, \partial _{i}ln|h|\right )\, \delta g_{bd} \sqrt{|h|} d^3 y= \int_{\Sigma_0} \triangledown_{3i}\left( [A^{ibd}]\,\delta g_{bd}\right) \, \sqrt{|h|}  \, d^3 y-\\- \int_{\Sigma_0}\,\left ( [\partial _{i}A^{ibd}] +[A^{ibd}]\, \Gamma _{ai}^{a}\right )\, \delta g_{bd} \sqrt{|h|} d^3 y,
\end{multline}
\par Similarly to the above reasoning for $U^n$, the term $\int_{\Sigma_0} \triangledown_{3i}\left( [A^{ibd}]\,\delta g_{bd}\right) \, \sqrt{| h|} \, d^3 y$ vanishes when applying the Stokes' theorem on the hypersurface. As a result we get:
$$
  [A^{abd}]\, \triangledown _{a}\delta g_{bd}=[A^{nbd}]\, \partial _{n}\delta g_{bd}- \left ( [\partial _{i}A^{ibd}]+[A^{ibd}]\, \Gamma _{ai}^{a}+[A^{acd}]\, \Gamma _{ac}^{b} +[A^{acb}]\, \Gamma _{ac}^{d}\right ) \, \delta g_{bd}\, .
$$
\par Taking into account all of the above, the equation  (\ref{22}) can be presented in the form:
\begin{multline} \label{30}
 \left\{ \right.\beta _{1}\left ( [\triangledown ^{n}R^{bd}]+[\partial _{k}\left ( g^{kn}R^{bd} \right )]+g^{kn}\Gamma _{ak}^{a}[R^{bd}] +g^{an}\Gamma _{ac}^{b}[R^{cd}]+g^{an}\Gamma _{ac}^{d}[R^{cb}]\right )+\\+\frac{1}{2}\, \beta _{2}\left (g^{bd}[\partial ^{n}R] +[\partial _{k}\left ( g^{kn} g^{bd}R\right )] +g^{kn}\, g^{bd}\Gamma _{ak}^{a}[R]+g^{an}g^{cd}\, \Gamma _{ac}^{b}[R] \right )-\\-\frac{1}{2}(\beta _{1}+\beta _{2})\left ( [\triangledown ^{b}R^{nd}]+[\partial _{k}\left ( g^{nb}R^{kd} \right )]+g^{nb} \, \Gamma _{ak}^{a}[R^{kd}]+g^{nd}\, \Gamma _{ac}^{b}[R^{ac}]\right )-\\-\frac{1}{2}(\beta _{1}+\beta _{2})\left ([\triangledown ^{d}R^{nb}]+[\partial _{k}\left ( g^{nd}R^{kb} \right )]+g^{nd} \, \Gamma _{ak}^{a}[R^{kb}]+g^{nb}\, \Gamma _{ac}^{d}[R^{ac}] \right ) +\\+\frac{1}{2}\, \beta _{2} \, g^{an}g^{cb}\, \Gamma _{ac}^{d}[R]-\frac{1}{2}(\beta _{1}+\beta _{2})\left(g^{nc}\, \Gamma _{ac}^{b}[R^{ad}]+g^{nc}\, \Gamma _{ac}^{d}[R^{ab}] \right) \left. \right \}\delta g_{bd}+\\+
\left \{\frac{1}{2} [R]\left (g^{bn} g^{dn}(\beta _{1}+\beta _{2})-g^{nn} g^{bd}\, \beta _{2}\right ) -\beta _{1}g^{nn}\, [R^{bd}]\right \}\partial _{n}\delta g_{bd}=\\=8\pi\,  S^{bd}\, \delta g_{bd}\,, \; k\neq n .
\end{multline}
It should be noted here that when deriving  (\ref{30}) the following consequence of  (\ref{24}-\ref{27}) was used:
\begin{equation} \label{29a}
  [R^{nb}]=\frac{1}{2} g^{nb}[R].
\end{equation}
\par It can be shown with the help of  (\ref{29a}) that the multiplier of $\triangledown_{n} \delta g_{bd}$ is equal to zero if at least one of the indexes $b$ or $d$ is equal to $n$:
$$
-\frac{1}{2}\beta_{2}\,  g^{nn} g^{bn} [R]+\frac{1}{2}(\beta _{1}+\beta _{2})g^{nn}g^{bn}[R]-\beta_{1}\, g^{nn}[R^{bn}]=0 \, .
$$
\par As noted in Ref.~\refcite{Berezin2020}, the variation $\delta (\partial_{n} g_{ij})$ of the derivatives of the metric on the hypersurface relates to the variation $\delta g_{ij}$ of the metric on the hypersurface but the connection between them is in a sense arbitrary, because the field equations in quadratic gravity are of the fourth order in terms of the derivatives of the metric tensor. It means their solutions are not uniquely determined by the initial conditions on the metric and its first derivatives. Therefore, we are forced to demand:
\begin{equation} \nonumber
    \delta (\partial_{n} g_{i j})= B^{k l }_{i j}(y) \, \delta g_{k l}, \quad i,j,l,k \neq n,
\end{equation}
where $B^{k l }_{i j}(y)$ are arbitrary functions. 
\par The appearance of arbitrary functions is indirectly connected to the presence of the derivative of the $\delta$-function in the equations of motion and serves as a marker of a double layer.
\par Eventually the field equations for a singular hypersurface of arbitrary type in quadratic gravity are obtained:
\begin{equation} \label{30a}
\beta _{2}\left ( \frac{\varepsilon }{2}\,  [\partial ^{n}R]-[\triangledown ^{n}R^{nn}]\right )+\frac{1}{2}\,(\beta _{1}+\beta _{2})\,\Gamma _{ij}^{n}\left ( g^{in} g^{jn}[R]-2\varepsilon [R^{ij}]\right )=8\pi\,  S^{nn}\, ,
\end{equation}
\begin{multline} \label{30b}
 \frac{1}{2}(\beta _{1}-\beta _{2})[\triangledown ^{n}R^{in}]+\beta _{1}g^{an}\Gamma _{ac}^{n}[R^{ci}]+\frac{1}{2}\, \beta _{2}\left \{g^{in}[\partial ^{n}R]  +g^{an}g^{ci}\, \Gamma _{ac}^{n}[R] \right \}-\\-\frac{1}{2}(\beta _{1}+\beta _{2})\left\{ \right.  -\frac{1}{2}g^{an}g^{cn}\, \Gamma _{ac}^{i}[R]-\frac{1}{2}g^{kn}\, g^{in}\Gamma _{ak}^{a}[R]+\varepsilon \, \Gamma _{ac}^{i}[R^{ac}]-\frac{1}{2}[\partial _{k}\left ( g^{kn} g^{in}R\right )]+\\+[\triangledown ^{i}R^{nn}]+\varepsilon [\partial _{k}R^{ki} ]+\varepsilon  \, \Gamma _{ak}^{a}[R^{ki}]+g^{ni}\, \Gamma _{ac}^{n}[R^{ac}]+g^{nc}\, \Gamma _{ac}^{n}[R^{ai}]  \left. \right \}=8\pi\,  S^{in},  
\end{multline}
\begin{multline} \label{30c}
  \beta _{1}\left \{ [\triangledown ^{n}R^{ij}]+[\partial _{k}\left ( g^{kn}R^{ij} \right )]+g^{kn}\Gamma _{ak}^{a}[R^{ij}] +g^{an}\Gamma _{ac}^{i}[R^{cj}]+g^{an}\Gamma _{ac}^{j}[R^{ci}]\right \}+\\+\frac{1}{2}\, \beta _{2}\left \{[\partial _{k}\left ( g^{kn} g^{ij}R\right )] +g^{kn}\, g^{ij}\Gamma _{ak}^{a}[R]+g^{an}g^{cj}\, \Gamma _{ac}^{i}[R]+g^{an}g^{ci}\, \Gamma _{ac}^{j}[R] \right \}-\\-\frac{1}{2}(\beta _{1}+\beta _{2})\left \{ [\partial _{k}\left ( g^{ni}R^{kj} \right )]+g^{ni} \, \Gamma _{ak}^{a}[R^{kj}]+g^{nj}\, \Gamma _{ac}^{i}[R^{ac}]+g^{nc}\, \Gamma _{ac}^{i}[R^{aj}]\right \}-\\-\frac{1}{2}(\beta _{1}+\beta _{2})\left \{[\partial _{k}\left ( g^{nj}R^{ki} \right )]+g^{nj} \, \Gamma _{ak}^{a}[R^{ki}]+g^{ni}\, \Gamma _{ac}^{j}[R^{ac}]+g^{nc}\, \Gamma _{ac}^{j}[R^{ai}] \right \}+\\+\frac{1}{2}\, \beta _{2} \, g^{ij}[\partial ^{n}R]-\frac{1}{2}(\beta _{1}+\beta _{2})\left\{[\triangledown ^{j}R^{ni}]+[\triangledown ^{i}R^{nj}] \right\}+\\
+\left \{\frac{1}{2} [R]\left (g^{kn} g^{ln}(\beta _{1}+\beta _{2})-\varepsilon\, g^{kl}\, \beta _{2}\right ) -\varepsilon \, \beta _{1}\, [R^{kl}]\right \}\, B_{kl}^{ij}(y)=8\pi\,  S^{ij}, 
\end{multline}
where $ i,j,l,k \neq n$ .
\par It follows from these equations that for $\beta_1=\beta_2=0$ there is no singular hypersurface. This combination corresponds to the quadratic Gauss-Bonnet term. It is known that in four dimensions it does not contribute to the equations of motion in the bulk, since it reduces to a total divergence of some vector \cite{Chern,Novikov,Yale}. As it turned out, the contribution to the motion equations on the $\Omega^{\pm}$ boundaries including $\Sigma_0$ as a part of $\partial \Omega^{\pm}$ is also zero if the Lichnerowicz conditions are satisfied. The absence of terms with $\alpha_4$ in the motion equations is also related to them.
\par For a lightlike double layer, the system of equations  (\ref{30a}-\ref{30c}) can be simplified using the fact that $g^{nn}|_{\Sigma_0}=\varepsilon=0$ for this case:
\begin{equation} \label{31a}
\frac{1}{2}\left(\beta _{1} -\beta _{2}\right ) \Gamma _{ij}^{n}\,  g^{in} g^{jn}[R]=\frac{1}{4}\left ( \beta _{1} -\beta _{2}\right ) g^{nk} g^{in} g^{jn}\, \partial _{k}g_{ij}\, [R]=0=8\pi S^{nn}, 
\end{equation}
\begin{multline} \label{31b}
 \frac{1}{2}(\beta _{1}-\beta _{2})g^{nk}[\triangledown_{k}R^{in}]+\beta _{1}g^{an}\Gamma _{ac}^{n}[R^{ci}]+\frac{1}{2}\, \beta _{2}\left \{g^{in}g^{nk}[\partial_{k}R]  +g^{an}g^{ci}\, \Gamma _{ac}^{n}[R] \right \}-\\-\frac{1}{2}(\beta _{1}+\beta _{2})\left \{-\frac{1}{2}g^{an}g^{cn}\, \Gamma _{ac}^{i}[R]-\frac{1}{2}g^{kn}\, g^{in}\Gamma _{ak}^{a}[R]-\frac{1}{2}[\partial _{k}\left ( g^{kn} g^{in}R\right )]\right \}-\\-\frac{1}{2}(\beta _{1}+\beta _{2})\left \{[\triangledown ^{i}R^{nn}]+g^{ni}\, \Gamma _{ac}^{n}[R^{ac}]+g^{nc}\, \Gamma _{ac}^{n}[R^{ai}] \right \}=8\pi\,  S^{in},  
\end{multline}
\begin{multline} \label{31c}
 \beta _{1}\left \{ g^{nk}[\triangledown_{k}R^{ij}]+[\partial _{k}\left ( g^{kn}R^{ij} \right )]+g^{kn}\Gamma _{ak}^{a}[R^{ij}] +g^{an}\Gamma _{ac}^{i}[R^{cj}]+g^{an}\Gamma _{ac}^{j}[R^{ci}]\right \}+\\+\frac{1}{2}\, \beta _{2}\left \{ [\partial _{k}\left ( g^{kn} g^{ij}R\right )] +g^{kn}\, g^{ij}\Gamma _{ak}^{a}[R]+g^{an}g^{cj}\, \Gamma _{ac}^{i}[R]+g^{an}g^{ci}\, \Gamma _{ac}^{j}[R] \right \}-\\-\frac{1}{2}(\beta _{1}+\beta _{2})\left\{ \right. [\partial _{k}\left ( g^{ni}R^{kj} \right )]+g^{ni} \, \Gamma _{ak}^{a}[R^{kj}]+g^{nj}\, \Gamma _{ac}^{i}[R^{ac}]+g^{nc}\, \Gamma _{ac}^{i}[R^{aj}]+\\+[\triangledown ^{i}R^{nj}]+[\triangledown ^{j}R^{ni}]+[\partial _{k}\left ( g^{nj}R^{ki} \right )]+g^{nj} \, \Gamma _{ak}^{a}[R^{ki}]+g^{ni}\, \Gamma _{ac}^{j}[R^{ac}]+g^{nc}\, \Gamma _{ac}^{j}[R^{ai}] \left. \right \}+\\+\frac{1}{2}\, \beta _{2}\,g^{ij}g^{nk}[\partial_{k}R]
+\frac{1}{2} [R]\, g^{kn} g^{ln}(\beta _{1}+\beta _{2})\, B_{kl}^{ij}(y)=8\pi\,  S^{ij},  \; i,j,k,l \neq n.
\end{multline}
\par Let's explain why the factor $g^{nk} g^{in} g^{jn}\, \partial _{k}g_{ij}$ in the equation (\ref{31a}) is zero for a null hypersurface:
\begin{multline} \nonumber
 g^{nk} g^{in} g^{jn}\, \partial _{k}g_{ij}=N^{a}N^{b}N^{c}\, \partial_{c}g_{ab}=\partial _{c}\left (N^{a}N^{b}N^{c}\, g_{ab}  \right )-2\, g_{ab}\, N^{b}\, N^{c}\, \partial _{c}N^{a}=\\=-2\, N_{a}\, N^{c}\, \partial _{c}\, N^{a}=-2\delta _{a}^{n}\, g^{nc}\, \partial _{c}\, g^{na}=0\, .
\end{multline}
It means that $S^{nn}$ is zero in any coordinate system, since it is a scalar: $S^{nn}=N_{a}N_{b}S^{ab}$.
\par Firstly it means that there is no “external pressure” for any null singular hypersurface. Secondly, lightlike double layer is possibly a source of radiation, since the “external flow” $S^{ni}=S^{i}_{a}\,N^{a}$ is presumably associated with radiation, and it cannot be zero if we are talking about a double layer and not a thin shell.
\par Without going into details, we note that the physical meaning of the surface energy-momentum tensor components for the null case can be associated with the thermodynamic interpretation of the gravitational moment on the lightlike hypersurface, which is presented in Refs.~\citen{Dey,Bhattacharya,Chakraborty}. In addition, it will be shown below that in the special coordinates $\{ n,\lambda,\theta^A \}$ the “external flow” has only one nonzero component which can be expressed in the invariant form:
$$
S^{n \lambda}= S^{ab} N_{a} l_{b}\, ,
$$
where $\lambda$ is a parameter varying along the lightlike geodesic congruence forming $\Sigma_0$, $l^{a}$ is an auxiliary null vector. Projections of this type in the aforementioned articles are associated with the analogue of the first law of thermodynamics for the geometric quantities describing generic null hypersurface.
\par During derivation of (\ref{30a}-\ref{30c}) the hypersurface was considered to be given a priori. An inverse problem arises in applications, when the solutions of the equations of motion in the domains $\Omega^{\pm}$ - $g^{\pm}_{ab}$ are known, and it is required to find the equation of the hypersurface on which the matching takes place. In this case, the equations with $S^{nn}$ and $S^{ni}$ in the right hand side together with the Lichnerowicz conditions define the singular hypersurface itself, while the rest of equations are necessary to find $B^{ij}_{kl}(y)$.
\par For thin shells the situation is different. The main criterion for the fact that $\Sigma_0$ is a thin shell and not a double layer is the absence of arbitrary functions in the motion equations. This condition is equivalent to the absence of jumps in the Ricci tensor for timelike and spacelike hypersurfaces: $[R^{bd}]=0$. However, it is a continuity of the scalar curvature: $[R]=0$ for a null thin shell. Moreover, some jumps in the Ricci tensor components can be nonzero, in particular, $[R_{nn}]$, since $g^{nn}|_{\Sigma_0}=0$ for a null hypersurface. The continuity of quantities corresponding to each type of hypersurface not only ensures the absence of arbitrary functions, but alsoprovides not only the absence of arbitrary functions, but also implies that $S^{nn}=S^{ni}=0$.
\par The motion equations for timelike and spacelike thin shells are obtained from (\ref{30a}-\ref{30c}) by imposing the additional condition $[R^{bd}]=0$:
\begin{multline} \label{30d}
S^{nn}=S^{ni}=0, \\
 \beta _{1} [\triangledown ^{n}R^{ij}]+\frac{1}{2}\, \beta _{2} \, g^{ij}[\partial ^{n}R] -\frac{1}{2}(\beta _{1}+\beta _{2})\left \{ [\triangledown ^{i}R^{nj}]+[\triangledown ^{j}R^{ni}]\right \}=8\pi\,  S^{ij} \, .
\end{multline}
\par Similarly, for lightlike thin shells we get:
\begin{multline}\label{30e}
S^{nn}=S^{ni}=0,\\
\beta _{1}\left \{ g^{nk}[\triangledown_{k}R^{ij}]+[\partial _{k}\left ( g^{kn}R^{ij} \right )]+g^{kn}\Gamma _{ak}^{a}[R^{ij}] +g^{an}\Gamma _{ac}^{i}[R^{cj}]+g^{an}\Gamma _{ac}^{j}[R^{ci}]\right \}-\\-\frac{1}{2}(\beta _{1}+\beta _{2})\left\{ \right. [\partial _{k}\left ( g^{ni}R^{kj} \right )]+g^{ni} \, \Gamma _{ak}^{a}[R^{kj}]+g^{nj}\, \Gamma _{ac}^{i}[R^{ac}]+g^{nc}\, \Gamma _{ac}^{i}[R^{aj}]+\\+[\triangledown ^{i}R^{nj}]+[\triangledown ^{j}R^{ni}]+[\partial _{k}\left ( g^{nj}R^{ki} \right )]+g^{nj} \, \Gamma _{ak}^{a}[R^{ki}]+g^{ni}\, \Gamma _{ac}^{j}[R^{ac}]+g^{nc}\, \Gamma _{ac}^{j}[R^{ai}] \left. \right \}=\\=8\pi\,  S^{ij},  \; i,j,k \neq n.
\end{multline}
\par In addition, it will be shown below that for a lightlike hypersurface the condition $[R]=0$ is satisfied in the case when $\Sigma_0$ is a leaf of some null foliation of $\Omega$.

\section{Junction conditions for null case}
\par As noted before the continuity of the metric on the singular hypersurface under consideration can be achieved by selecting the appropriate coordinates. For timelike and spacelike hypersurfaces it is usually Gaussian normal coordinates.
\par For lightlike hypersurfaces, there are some analogs of Gaussian normal coordinates, such as Gaussian null coordinates (GNC) \cite{Moncrief,Friedrich,Racz} or coordinates of null surface foliation (NSFC) \cite{Parattu,Jezierski,Padmanabhan}. This article, however, has taken a more minimalistic approach. It does not require additional information about the behavior of a normal vector field outside of $\Sigma_0$. Coordinates which are a special case of $\{n,y^i\}$ for a null hypersurface were constructed with the help of the formalism described in the book Ref.~\refcite{Poisson} by E. Poisson. A short summary of this process is presented below.
\par Let us consider the lightlike hypersurface $\Sigma_0$ defined in $\Omega^{\pm}$ domains with arbitrary coordinates $\{x^{\pm}\}$ by the equations: $n^{\pm}(x^{\pm })=0$. A normal vector field to a hypersurface is defined as: $N_{\pm a}=\partial_{a}n^{\pm}(x^{\pm})$. Let us temporarily omit the notation $\pm$ for convenience and write down all the equations in one of the regions.
\par Since the scalar $N^{a} N_{a}$ is zero on $\Sigma_0$ for any null hypersurface, the normal vector $N^{a}$ is defined up to multiplication by an arbitrary scalar function and also tangent to the $\Sigma_0$.
\par It can be demonstrated that $N^{a}$ is tangent to the lightlike geodesics generators of $\Sigma_0$ :
\begin{equation} \nonumber
\triangledown_b N_{a} \,  N^{ b}=\triangledown_{ab} n \, \partial^{b} n =\triangledown_{b a} n \, \partial^{b} n =\frac{1}{2}\triangledown_a \left(N_b \,  N^b \right) .    
\end{equation}
\par Since $N^{a} N_{a}$ is zero on hypersurface its gradient is directed along $N_{a}$, then
\begin{equation} \nonumber
\triangledown_b N_{a} \, N^{ b}=\kappa N_{a},    
\end{equation}
where $\kappa$ is some scalar. 
\par Let's choose an arbitrary (not necessarily affine) parameter $\lambda$ on the above-mentioned null hypersurface generators and two additional coordinates $\theta^{A}, A=2,3$ for labeling geodesics. Together they form the internal coordinate system $y^{ i}=\{\lambda,\theta^{ A}\}$ of the hypersurface. The $\lambda$ is affine if the equation $n(x)=const$ is a leaf of some null foliation of $\Omega$.
\par The induced metric on $\Sigma_0$ is:
\begin{equation} \nonumber
  ds^{2}_{\Sigma_0}=g_{ab}dx^{a} (y^{i}) dx^{ b}(y^{ j})=g_{ab}e^{a}_{i}e^{ b}_{j}dy^{i} dy^{j},\quad e^{a}_{i}=\frac{\partial x^{a}}{\partial y^{i}}.
\end{equation}
\par The vector fields $e^{a}_{i}$ are orthogonal to the normal since they are tangent to curves lying on the $\Sigma_0$:
\begin{equation} \nonumber
N_{a} e^{ a}_{A}=0,\quad N_a e^{ a}_{1}=N_a \; \frac{\partial x^{a}}{\partial \lambda}=N_{a} N^{a}=0.    
\end{equation}
\par For null hypersurfaces the induced metric is effectively two-dimensional:
\begin{equation} \nonumber
  ds^{2}_{\Sigma_0}=g_{ab}e^{a}_{A}e^{ b}_{B}d \theta^{A} d \theta^{ B}=\sigma_{AB}d \theta^{ A} d \theta^{ B} .
\end{equation}
\par The system of vector fields $\{ N^{a},e^{a}_{A} \}$ can be extended to a basis in the restriction of the manifold $\Omega$ vector bundle to the hypersurface - $T\Omega |_{\Sigma_0} $ if we find an auxiliary null vector field $l^{a}$ with the following properties:
\begin{equation} \nonumber
l^{ a} N_{a}=1, \quad l_{ a} e^{a}_{A}=0, \quad l^{a} l_{ a}=0.    
\end{equation}
\par The existence of such vector field as well as the completeness of the system $\{ l^{a},N^{a},e^{a}_{A} \}$ in $T\Omega |_{\Sigma_0}$ are demonstrated in particular in the Ref.~\refcite{Poisson} and Ref.~\refcite{SenovillaMars}. The latter also introduces the concept of a rigging vector, which is a generalization of an auxiliary vector $l^{a}$ for singular hypersurfaces of a mixed causal nature.
\par The basis $\{ l^{a},N^{a},e^{a}_{A} \}$ can be used to write down the completeness relations for the inverse metric:
\begin{equation} \label{31}
    g^{ab}=l^{a} N^{ b}+l^{b} N^{a}+\sigma^{ A B} e^{a}_{A} e^{ b}_{B} .
\end{equation}
\par Choosing the $\{n,\lambda,\theta^{ A}\}$ as local coordinates we obtain:
\begin{equation} \label{32}
    N_a=\delta^{n}_{a}, \quad N^{a}=\frac{\partial x^{a}(y^{i})}{\partial \lambda}=\delta^{a}_{\lambda},\quad l_{\lambda}=l^{n}=1, \quad l_{n}=-l^{\lambda}, \quad l_{A}=0.
\end{equation}
\par From (\ref{31}) follows that:
\begin{equation} \label{33}
    g^{ n \lambda }=1, \quad  g^{ n n}=g^{ n A}=0, \quad g^{ \lambda \lambda }=2 \; l^{\lambda}, \quad g^{\lambda A }= l^{A}, \quad g^{ A B }= \sigma^{ A B},
\end{equation}
\begin{multline} \label{34}
    g_{ n \lambda }=1, \quad  g_{ \lambda \lambda}=g_{ \lambda A}=0, \quad g_{ n n }=-2 \; l^{\lambda}+ \sigma_{A B} l^{B} l^{A},\\  g_{n A }= -\sigma_{A B} l^{B}, \quad g_{ A B }= \sigma_{A B}.
\end{multline}
\par The form of the metric described by (\ref{33},\ref{34}) is valid only on $\Sigma_0$, i.e. for $n=0$. The second derivatives with respect to $n$ of $g_{ n \lambda }, g_{ \lambda \lambda}, g_{ \lambda A}$ and their jumps on $\Sigma_0$ are generally nonzero. The continuity of the first derivatives of the metric with respect to $n$ is ensured by the Lichnerowicz conditions. Relations  (\ref{33},\ref{34}) extend to the neighborhood of $\Sigma_0$ only if the vector field $\partial_a n(x)$ is null also in some neighborhood of $\Sigma_0$ but even in this case: $\partial^2_{nn} \; g_{\lambda n}\neq 0$.
\par A proper coordinates common for $\Omega^{\pm}$ are constructed using the following procedure. The coordinates $n^{+}$ and $n^{-}$ are continuously merged into one coordinate $n$ and either $y^{+i}$ or $y^{-i}$ serve as remaining coordinates. The continuity of the metric on the hypersurface is provided by (\ref{33}) and the appropriate choice of the functions $y^{+i}(y^{-j})$.
\par Let us use the coordinates $\{ n,\lambda,\theta^{ A} \}$ constructed above to simplify the equations  (\ref{31a}-\ref{31c},\ref{30e}). In order to do this let's write down the jumps they contain:
\begin{equation} \label{42}
    [R]=2 \; [R_{\: nn}^{n\: \:\:  n}]=[\partial^{2}_{n n} g_{\lambda \lambda}], \quad [\partial_i R]=[\partial_i \partial^{2}_{n n} g_{\lambda \lambda}],
\end{equation}
\begin{equation} \label{43}
 [R^{ab}]=\frac{1}{2}\left ( \delta_{\lambda }^{a} \, g^{bc}\, [\partial _{nn}^{2}g_{c\lambda }] +\delta_{\lambda }^{b} \, g^{ac}\, [\partial _{nn}^{2}g_{c\lambda }]-\delta_{\lambda }^{a} \,\delta_{\lambda }^{b} \, g^{cd}\, [\partial _{nn}^{2}g_{c d }]  \right ) ,
\end{equation}
\begin{multline} \label{44}
  [\triangledown _{\lambda }R^{bd}]=\partial _{\lambda }[R^{bd}]+\delta _{\lambda }^{b }\, \left ( \Gamma _{\lambda \lambda }^{\lambda } [R^{\lambda  d}]+\Gamma _{\lambda  A}^{d}[R^{\lambda A}]+\frac{1}{2}\, \Gamma _{\lambda 
 n}^{d}\, [R]\right )+\\+\delta _{\lambda }^{d }\, \left ( \Gamma _{\lambda \lambda }^{\lambda } [R^{\lambda  b}]+\Gamma _{\lambda  A}^{b}[R^{\lambda A}]+\frac{1}{2}\, \Gamma _{\lambda 
 n}^{b}\, [R]\right ),
\end{multline}
\begin{equation} \label{45}
  [\triangledown _{i} R^{n d}]=\frac{1}{2}\, \delta _{\lambda }^{d}\left ( \partial _{i}[R]+\Gamma _{i n}^{n}\, [R] +2\, \Gamma _{i  A}^{n}\, [R^{\lambda A}]\right )+\frac{1}{2}\, \Gamma _{i \lambda }^{d}\, [R] .
\end{equation}
 \par Thus, we obtain the equations for the null double layer in quadratic gravity in the coordinates $ \{ n, \lambda, \theta^{A} \} $:
 \begin{equation} \label{47a}
     S^{nn}=0,
 \end{equation}
\begin{multline} \label{47}
\frac{1}{2}\, \beta _{2}\left \{g^{\lambda \lambda }[\partial_{\lambda }R] +[\partial _{\lambda }\left (g^{\lambda \lambda }R\right )] + g^{\lambda \lambda }\Gamma _{a\lambda }^{a}[R]+2g^{c\lambda }\, \Gamma _{\lambda c}^{\lambda }[R]\right \}-\\-(\beta _{1}+\beta _{2})\left \{\frac{1}{2}[\partial ^{\lambda }R]+\frac{1}{2}g^{\lambda k}[\partial _{k}R]+\frac{1}{4}g^{\lambda A}g^{\lambda B}\, \partial _{\lambda }\sigma _{AB}\, [R]-\frac{1}{2}\Gamma _{An}^{A}[R]+\frac{1}{2}\Gamma _{\lambda n}^{\lambda }[R]\right \}+\\+\beta _{1}\left \{2 [\triangledown_{\lambda }R^{\lambda \lambda }]+\Gamma _{a\lambda }^{a}[R^{\lambda \lambda }] \right \}+\frac{1}{2} [R]\,(\beta _{1}+\beta _{2})\, B_{\lambda \lambda }^{\lambda \lambda }(y)=8\pi\,  S^{\lambda \lambda },
\end{multline}
\begin{equation} \label{48}
(\beta _{1}+\beta _{2})\, \left (\,  [\partial _{\lambda }R]+ \Gamma _{k\lambda }^{k}\, [R]\,  \right )=16\pi\,  S^{n\lambda},  
\end{equation}
\begin{multline} \label{49}
\frac{1}{2}\beta _{2}\left \{2g^{\lambda A}[\partial _{\lambda }R] +\partial _{\lambda }g^{\lambda A}[R]+g^{\lambda A}\Gamma _{a\lambda }^{a}[R]+g^{kA}\Gamma _{\lambda k}^{\lambda }[R] \right \}-\\-\frac{1}{2}\beta_{1}\,g^{a\lambda }\Gamma _{a\lambda }^{A}[R]+\beta_{1}\left \{ 2[\triangledown _{\lambda }R^{\lambda A}] +\Gamma _{a\lambda }^{a}[R^{\lambda A}]\right \}-\\-\frac{1}{2}(\beta _{1}+\beta _{2})\left \{ [\partial ^{A}R] -B_{\lambda \lambda }^{\lambda A}(y)[R]\right \}=8\pi S^{\lambda A}\, ,
\end{multline}
\begin{multline} \label{50}
\frac{1}{2}\beta _{2}\left \{ 2\sigma ^{AB}\, [\partial _{\lambda }R] +\partial _{\lambda }\sigma ^{AB}\, [R]+\sigma ^{AB}\, \Gamma _{a\lambda }^{a}[R]\right \}+\\+\frac{1}{4}(\beta _{2}-\beta _{1})\, \sigma ^{AC}\sigma ^{BD}\, \partial _{\lambda }\sigma _{CD}[R]+\frac{1}{2}(\beta _{2}+\beta _{1})B_{\lambda \lambda }^{AB}(y)[R]=8\pi S^{AB}\, .
\end{multline}
\par In addition, it is necessary to pay attention to an important special case when $\Sigma_0$ at is a leaf of some null foliation. It means that vector field $\partial_a n(x)$ is null also in some neighborhood of $\Sigma_0$ and from  (\ref{42}-\ref{45}) we get:
\begin{equation} \label{36}
    [R]=[R^{\lambda A}]=[\partial_i R]=[\triangledown_i R^{nd}]=0,
\end{equation}
\begin{equation} \label{51}
 [\triangledown_{\lambda} R^{b d}]=\delta _{\lambda }^{b}\, \delta _{\lambda }^{d} \, [\partial_{\lambda} R^{\lambda \lambda}] .
\end{equation}
\par Thus, in this case singular hypersurface is a thin shell, the only non-zero motion equation for which in the coordinates $ \{ n, \lambda, \theta^{A} \} $ is:
\begin{equation} \label{37}
 \beta _{1}\left \{2\,  [\partial _{\lambda }R^{\lambda \lambda }]+\Gamma _{a\lambda }^{a}\, [R^{\lambda \lambda }] \right \}-\frac{1}{2}(\beta _{1}+\beta _{2})\, [\partial _{n}R]=8\pi\,  S^{\lambda \lambda } \, . 
\end{equation} 

\section{Spherically symmetrical null thin shells}

This section explores spherically symmetric lightlike singular hypersurfaces separating two spherically symmetric spacetimes.
\par Let's consider a spherically symmetric metric of the most general form:
\begin{multline}\label{52}
     ds^{2}=\gamma _{\alpha \beta }(x)\, dx^{\alpha }dx^{\beta}-r^2(x)\, \left ( d\theta ^{2}+\sin^{2 }\theta \,  d\phi ^{2}\right )=\\=r^2(x)\left ( \widetilde{\gamma }_{\alpha \beta }(x) \, dx^{\alpha }dx^{\beta}- d\Omega^{2} \right )=r^2(x)\left (d\widetilde{s}^{2}_{2}- d\Omega ^{2} \right ) .
\end{multline}
Hereinafter $2 + 2$ decomposition of the metric in the case of spherical symmetry is used, the Greek indices take values
$\{0,1\}$. These indices apply to quantities related to the two-dimensional “metric” $\widetilde{\gamma}_{\alpha \beta}$.
\par For this type of spacetime the following relations hold:
\begin{equation}  \label{51a}
    R_{\theta \theta }=1+r \sigma +\Delta ,\quad R_{\phi \phi }=R_{\theta \theta } \; sin^{2} \theta,
\end{equation}
\begin{multline} \label{51b}
R_{\alpha \beta }=\frac{\gamma _{\alpha \beta }}{r^{2}}\left ( \frac{1}{2} \, \widetilde{R}+\Delta -r \, \sigma  \right )-\frac{2}{r} \, \triangledown _{\alpha }\triangledown _{\beta }\, r=\\=\widetilde{\gamma }_{\alpha \beta }\left ( \frac{1}{2} \, \widetilde{R}-\frac{\widetilde{\Delta }}{r^{2}} -\frac{\widetilde{\sigma }}{r} \right )-\frac{2}{r} \, \widetilde{\triangledown }_{\alpha }\widetilde{\triangledown }_{\beta }\, r+\frac{4}{r^{2}}\partial _{\alpha }r\, \partial _{\beta }r,
\end{multline}
\begin{equation} \label{51c}
    R=\frac{1}{r^{2}}\left ( \widetilde{R}-2 -6\,  r \, \sigma  \right ) .
\end{equation}
Here $\Delta$, $\sigma$, $\widetilde{R}$ are spherical geometry invariants:
\begin{multline} \label{51d}
    \Delta =g^{a b}\,  \partial _{a}r \, \partial_{b } r=\gamma ^{\alpha \beta }\,  \partial _{\alpha }r \, \partial _{\beta } r=\frac{1}{r^{2}}\, \widetilde{\Delta }, \\ \sigma =\square \, r -\frac{2}{r} \, \Delta=\gamma^{\alpha \beta}\, \triangledown_{\alpha } \triangledown_{\beta } \, r\,=\frac{1}{r^{2}} \, \widetilde{\sigma }=\frac{1}{r^{2}}\, \widetilde{\square }\, r,
\end{multline}
\begin{multline} \label{51e}
    \widetilde{R}=\frac{1}{\widetilde{\gamma }}\left ( -\partial_{11}^{2}\widetilde{\gamma}_{00}+2 \partial_{01}^{2}\widetilde{\gamma}_{01}-\partial_{00}^{2}\widetilde{\gamma}_{11}\right )+\frac{1}{2\widetilde{\gamma }^{2}}\left ( \widetilde{\gamma }_{11}\left ( \partial_{1}\widetilde{\gamma }_{00}\right )^{2}+\widetilde{\gamma }_{00}\left ( \partial_{0}\widetilde{\gamma }_{11}\right )^{2} \right )+\\+\frac{1}{2\widetilde{\gamma }^{2}}\left (\partial_{1}\widetilde{\gamma }_{00}\left ( -\partial_{0}\widetilde{\gamma}_{11}\, \widetilde{\gamma }_{10}+\partial_{1}\widetilde{\gamma}_{11}\, \widetilde{\gamma }_{00}-2\partial_{1}\widetilde{\gamma}_{01}\, \widetilde{\gamma }_{10}\right )\right)+\\+\frac{1}{2\widetilde{\gamma }^{2}}\left( \partial_{0}\widetilde{\gamma }_{11}\left ( \partial_{0}\widetilde{\gamma}_{00}\, \widetilde{\gamma }_{11} -2\, \partial_{0}\widetilde{\gamma}_{10}\, \widetilde{\gamma }_{10}\right )\right )+\\+\frac{1}{2\widetilde{\gamma }^{2}}\left ( \partial_{0}\widetilde{\gamma }_{00}\left ( \partial_{1}\widetilde{\gamma}_{11}\, \widetilde{\gamma }_{01} -2\, \partial_{1}\widetilde{\gamma}_{10}\, \widetilde{\gamma }_{11}\right )-2\, \partial_{0}\widetilde{\gamma }_{10}\left ( \partial_{1}\widetilde{\gamma}_{11}\, \widetilde{\gamma }_{00} -2\, \partial_{1}\widetilde{\gamma}_{10}\, \widetilde{\gamma }_{10} \right ) \right ).
\end{multline}
\par The notation $\widetilde{R}$, $\widetilde{\gamma}$, $\widetilde{\triangledown}$, $\widetilde{\square}$ is used for the two-dimensional scalar curvature, determinant, covariant derivative and Laplacian of the two-dimensional “metric” $\widetilde{\gamma}_{\alpha \beta}$ respectively.

\par For any spherically symmetric lightlike hypersurface in the geometry (\ref{51a}) given by the equation $n(x)=0$, the following is true directly on the hypersurface:
$$
\gamma_{00}\left ( \partial^{0}n(x) \right )^{2}+2\gamma_{01} \, \partial^{0}n(x)\,  \partial^{0}n(x) +\gamma_{11}\left ( \partial^{1}n(x) \right )^{2}=0 ,
$$
but if this equation can be solved with respect to the variable $\frac{\partial^{0}n(x)}{\partial^{1}n(x)}$ or its reciprocal for any values of $x$, then the function $n(x)$ defines a whole family of null hypersurfaces $n(x)=const$; i.e. the vector field $\partial^{\alpha} n(x)$ is null in the entire spacetime under consideration. Since it is a quadratic equation with respect to the specified variable, in order for it to always have at least one solution, $-\gamma$ must be non-negative. This condition is always satisfied for any Lorentzian spherically symmetric metric.
\par Thus, it was shown that for a lightlike spherically symmetric singular hypersurface in spherically symmetric geometry $[R]=0$ if the jumps are calculated starting only from the second derivatives due to the Lichnerowicz  conditions. Another consequence is that the metric  (\ref{52}) in the neighborhood of null spherically symmetric hypersurface in the coordinates $\{ n,\lambda,\theta,\phi\}$ has a special form:
\begin{equation} \label{56}
    ds^2=\gamma_{nn}\, dn^{2}+2\gamma_{n \lambda }\, dn\,  d\lambda -r^{2}(n,\lambda )\, d\Omega ^{2} \, .
\end{equation}
\par Let us express the jump $[\partial_{\lambda }R^{\lambda \lambda }]$ in the equation  (\ref{37}) in terms of spherical geometry invariants presented above:
\begin{equation} \label{53}
\Gamma _{a\lambda }^{a}\, [R^{\lambda \lambda }]=-\frac{2}{r^{2}}[\partial _{n}\Delta ],\quad[\partial_{\lambda }R^{\lambda \lambda }]=\partial _{\lambda }\left ( -\frac{2}{r} [\partial^{2}_{nn}r]\right )=\frac{1}{r^{2}}[\partial _{n}\Delta ]-\frac{1}{r}[\partial _{n}\sigma ] \, .
\end{equation}
Here we have used the relations (\ref{51b}-\ref{51d}) for the metric (\ref{56}) and also the fact that $\gamma_{n \lambda}|_{\Sigma_0}=1$:
$$
\sigma =\frac{1}{\gamma _{n\lambda }}\, \partial _{\lambda }\left ( 2\, \partial _{n}r-\frac{\gamma _{nn}}{\gamma _{n\lambda }}\, \partial _{\lambda }r \right ), \quad \Delta =-\frac{\gamma _{nn}}{\gamma _{n\lambda }}\, \left ( \partial _{\lambda } r\right )^{2}+\frac{2}{\gamma _{n\lambda }}\, \partial _{n}r\, \partial _{\lambda }r\, ,
$$
\begin{multline} \nonumber
 R_{nn}=\frac{\gamma _{nn}}{r^{2}}\left ( \frac{1}{2} \, \widetilde{R}+\Delta -r \, \sigma  \right )-\frac{2}{r}\, \partial _{nn}^{2}r+\frac{\partial_{n}r}{ \gamma _{n\lambda }\, r} \left (2\, \partial _{n}\gamma_{n\lambda }-\partial _{\lambda }\gamma _{nn}\right )+\\+\frac{\partial_{\lambda }r}{\gamma_{n\lambda } \, r} \left (  \partial _{n} \gamma _{nn}-2\, \frac{\gamma _{nn}}{\gamma _{n\lambda }}\, \partial _{n}\gamma_{n\lambda }+\frac{\gamma _{nn}}{\gamma _{n\lambda }} \, \partial _{\lambda }\gamma _{nn}\right )\, .  
\end{multline}
\par By means of (\ref{51c}) and (\ref{53}) we obtain the motion equation of a thin lightlike shell in the spherically symmetric case:
\begin{equation} \label{54}
-\frac{1}{2}(\beta _{1}+\beta _{2}) [\partial_{n}\widetilde{R}]+ r\,  (\beta _{1}+3\beta _{2})[\partial _{n} \sigma ]= 8  \pi \, r^{2}\, S^{\lambda}_{n}\, .
\end{equation}
It can be presented in the invariant form by multiplying both sides of the equation by $\partial_{\lambda}r$:
\begin{equation} \label{55}
-\frac{1}{2}(\beta _{1}+\beta _{2})\,  [g^{ab}\partial_{a}r \,\partial_{b}\widetilde{R}]+r\,  (\beta _{1}+3\, \beta _{2})[g^{ab}\partial_{b}r \,\partial _{a} \sigma ]= 8  \pi \, r^{2}\, S^{a}_{n} \, \partial_{a}r\, .
\end{equation}
\par The Lichnerowicz conditions can also be expressed in terms of spherical geometry invariants:
\begin{equation} \label{68}
  [\Delta]=0, \quad [\widetilde{R}]=0 ,
\end{equation}
where we have used the definition of $\widetilde{R}$ for the case of metric (\ref{56}):
$$
\widetilde{R}=\frac{1}{\widetilde{\gamma }_{n\lambda }}\, \partial _{\lambda }\left ( \frac{\partial _{\lambda }\widetilde{\gamma }_{nn}-2\, \partial _{n}\widetilde{\gamma }_{n\lambda }}{\widetilde{\gamma }_{n\lambda }} \right )\, .
$$
\par The relations  (\ref{68}) ensure that there are no jumps in $\partial_n \gamma_{n \lambda}$ and $\partial_n r$
while for the metric of the form  (\ref{56}) it is also necessary to require: $[\partial_n \gamma_{nn}]=0$. However, in the case of spherical symmetry, this component of the metric can always be eliminated by coordinate transformations.
\par For null singular hypersurfaces the general form of the hypersurface equations in the $\Omega^{\pm}$ domains is determined by the condition of zero interval. At the same time, the Lichnerowicz conditions (\ref{68}) together with  (\ref{54}) impose a certain restrictions on the metrics to be matched and on the surface energy-momentum tensor. Let us explain this fact with some examples. 
\par Any spherically symmetric metric can be reduced to the form:
\begin{equation} \nonumber
     ds^{2}=2 H(u,v) du dv - r^2 (u,v) d \Omega^2 .
\end{equation}
\par Since it is preserved under certain coordinate transformations, namely: $u\rightarrow \widetilde{u}(u),\, v\rightarrow \widetilde{v}(v)$ or $u \rightarrow v,\, v\rightarrow u$, without loss of generality we can assume that the hypersurface in the domains $\Omega^{\pm}$ is given by the equations: $u^{\pm}=0=n^{\pm}$. Therefore, the coordinates $\lambda^{\pm}$ are defined by the relations: $d \lambda^{\pm}= H^{\pm}(o,v^{\pm})dv^{\pm}$. The function $\lambda^{+}(\lambda^{-})$ must be defined in such a way that the metric is continuous on the hypersurface. Since $\gamma^{\pm }_{n\lambda }|\Sigma _{ 0}=\frac{H^{\pm}(n,\lambda^{\pm})}{H^{\pm}(0,\lambda^{\pm})}|\Sigma _{0} =1$, only one condition remains: $r^{+}(0,\lambda^{+})=r^{-}(0,\lambda^{-})$. It also determines the relationship between the original coordinates $v^{+}(v^{-})$ on $\Sigma_0$. 
\par The Lichnerowicz conditions impose additional restrictions on the functions $r^{\pm}$ and $H^{\pm}$ on the hypersurface:
$$
\partial_{u^{+}} r^{+}(0,v^{+}(v^{-}))=\partial_{u^{-}} r^{-}(0,v^{-}), \quad \partial_{u^{+}} H^{+}(0,v^{+}(v^{-}))=\partial_{u^{-}} H^{-}(0,v^{-}) \, .
$$
\par In some cases the invariant form of (\ref{68}) allows us to determine whether the matching of $\Omega^{\pm}$ under consideration with a null singular hypersurface exist without resorting to special coordinates $\{n,\lambda\}$.
\par For example, for the class of metrics: $f(r) \, dt^2-f^{-1}(r)\, dr^2-r^2 d \Omega^2$, it follows from the continuity of $\Delta=-f(r)$ and $r$ on the hypersurface that $r=r_0=const$ on $\Sigma_0$. On the other hand, if $\Sigma_0$ is a lightlike hypersurface, then $f^{\pm}(r_0)=0$. In this case $[\Delta]=0$ means that $f^{+}(r)= f^{-}(r)$, i.e. $\Omega^{\pm}$ are the same and matching doesn't exist. In particular, it means that, unlike the general relativity \cite{Poisson}, in quadratic gravity there is no null thin shell separating the Schwarzschild spacetime and the Minkowski vacuum.
\par However, it will be shown below that for spherically symmetric lightlike singular hypersurfaces in quadratic gravity the Lichnerowicz conditions are not necessary. In this situation, the matching between Schwarzschild geometry and Minkowski vacuum with the null hypersurface is actually possible, but it is a null double layer, not a thin shell.

\section{Conformal gravity}

\par The following values of the coefficients in the original quadratic gravity Lagrangian: 
$$
\alpha_2=-2\alpha_1, \quad \alpha_3=\frac{1}{3}\alpha_1,\quad \alpha_4=\alpha_5=0 ,
$$
lead to the conformal gravity Lagrangian: 
\begin{multline} \label{69}
  S_{q} =-\frac{1}{16 \pi} \int_{\Omega} \sqrt{-g} \,  \alpha_1 \left(R_{abcd} \, R^{abcd}-2 R_{ab}\, R^{ab}+\frac{1}{3} R^2\right)\, d^{4}x=\\ =-\frac{1}{16 \pi} \int_{\Omega} \sqrt{-g} \, \alpha_1 \, C^2 \, d^{4}x .
\end{multline}
\par The corresponding equations of motion were derived for the first time by R. Bach \cite {Bach}. They are a special case of (\ref {18}):
\begin{equation} \label{70}
    \triangledown_{c } \triangledown _{d }\, C^{a c b d}+\frac{1}{2} \, C^{a c b d } \,R_{c d }=\frac{2 \pi}{ \alpha_{1}}\, T^{a b} .
\end{equation}
Here, the sign in front of $\frac{1}{2}$ in the Bach equation is due to the chosen signature.
\par Certain types of spherically symmetric solutions of  (\ref {70}) were obtained in the article Ref.~\refcite{Berezin2016}. Since the radius acts as a conformal factor their general form coincides with (\ref {52}) provided that $r^2(x)$ are arbitrary functions.
\par The null thin shells separating two spherically symmetric solutions of (\ref {70}) will be considered next. In particular, we are interested in vacuum and Vaidya-type solutions.
\par The motion equation of null thin shell in spherically  symmetric conformal gravity is a special case of  (\ref{54}):
\begin{equation} \label{77}
     [\partial_{n}\widetilde{R}]=-\frac{12 \pi}{\alpha_1}  \, r^{2}\, S^{\lambda}_{n}\, .
\end{equation}

\subsection{Vacuum solutions}
\par Let's consider first the situation where $\Omega^{\pm}$ are two spherically symmetric vacuums of conformal gravity. 
\par The article Ref.~\refcite{Berezin2016} presented three types of  spherically symmetric conformal gravity vacua: two with constant two-dimensional scalar curvature $\widetilde{R}=\mp 2$ and one with variable $\widetilde{R}$. The two-dimensional “metrics” in double null coordinates for different vacua are:
\begin{equation} \label{72}
d\widetilde{s}^{2}_{2}=2 \widetilde{H}(u,v) \,du dv ,
\end{equation}
\begin{equation} \label{78}
    \widetilde{H}=\frac{2}{(u\pm v)^2}, \quad \widetilde{R}=\mp 2.
\end{equation}
\begin{multline} \label{79}
    \widetilde{H}=\frac{1}{2}|A(\widetilde{R}(u,v))|, \quad A(\widetilde{R})=\frac{1}{6}\left ( \widetilde{R}^{3}-12\widetilde{R}+C_0 \right ) ,\\
  \widetilde{R}=\widetilde{R}(v-u): \;\int \frac{d\widetilde{R}}{A(\widetilde{R})}=\frac{1}{2}(v-u), \; A(\widetilde{R})>0 ,\\
  \widetilde{R}=\widetilde{R}(v+u): \;\int \frac{d\widetilde{R}}{A(\widetilde{R})}=\frac{1}{2}(v+u), \; A(\widetilde{R})<0 ,\\
\end{multline}
where $C_0$ is some constant.
\par Condition $ [\widetilde{R}] = 0$ dictates the existence of the following combinations: matching a vacuum with a constant $ \widetilde {R} = \pm 2 $ and a vacuum with a variable $ \widetilde{R}$ and matching two vacua with variable $ \widetilde{R} $. Matching two vacua with coinciding constant $ \widetilde{R} $ is possible but according to the equation (\ref{77}) there is no thin shell in this case.
 \par Let's consider the case when $\Omega^{-}$ is a vacuum with a constant $\widetilde{R}=\pm 2$ and $\Omega^{+}$ is a vacuum with a variable $\widetilde{R}$.
 \par From the continuity of the two-dimensional scalar curvature, we obtain that the hypersurface equation in the $\Omega^{+}$ is $n^{+}(x^{+})=\widetilde{R}^{+}\mp 2=0$. Such a hypersurface could be null only if $C_0=\pm 16$ when $n^{+}=0$ is an analogue of the double horizon with respect to the variable $\widetilde{R}^{+}$. In the coordinates $\{n^{+}, u^{+}\}$ the two-dimensional “metric” in the $\Omega^{+}$ is:
 \begin{equation} \label{81}
    d\widetilde{s}_{2}^{+2}=\frac{1}{6}(n^{+})^2 (n^{+}\pm 6)du^{+2}+2du^{+}d n^{+}.
\end{equation}
\par In the domain $\Omega^{-}$ the hypersurface equation is $n^{-}(x^{-})= F^{-}(v^{-})=0$, where $F^{-}(v^{-})$ is an arbitrary function of $v^{-}$ (the case when $ F^{-} $ is an arbitrary function of $ u^{-} $ is similar to this one). In the coordinates $\{n^{-}, u^{-}\}$ the two-dimensional “metric” in the $\Omega^{-}$ is: 
\begin{equation} \label{81a}
    d\widetilde{s}_{2}^{-2}=\frac{4}{(u^{-}\mp f^{-}(n^{-}))^2}\frac{df^{-}}{dn^{-}}dn^{-}du^{-},
\end{equation}
where $f^{-}(n^{-})$ is a function inverse to $F^{-}(v^{-})$. 
\par From the continuity of the metric on $\Sigma_0$ it follows that:
$$
r^{+}(0,u^{+})=r^{-}(0,u^{-}) .
$$
\par The condition $[\Delta]=0$ defines the relationship between functions $r^{\pm}(x^{\pm})$ on the hypersurface:
\begin{equation} \label{82}
    \partial_{n^{+}}r^{+}|_{\Sigma_0}=\partial_{\widetilde{R}^{+}}r^{+}(2,u^{+})=\partial_{n^{-}} r^{-}|_{\Sigma_0}=\partial_{n^{-}}f^{-}(0)\, \partial_{v^{-}} r^{-}(f^{-}(0),u^{-}) .
\end{equation}
\par Let's consider next the matching of two vacua with variable $\widetilde{R}$. Without loss of generality we can assume that $\Sigma_0$ is described by the equations: $n^{\pm}(x^{\pm})=F^{\pm}(u^{\pm})=0$ in $\Omega^{\pm}$, where $F^{\pm}(u^{\pm})$ are arbitrary functions of variables $u^{\pm}$. The coordinates ${n^{\pm},\widetilde{R}^{\pm}}$ are the most suitable in this case:
\begin{equation} \label{83}
    d\widetilde{s}_{2}^{\pm 2}=A^{\pm} \left(\frac{df^{\pm}}{dn^{\pm}} \right)^{2} dn^{\pm 2}+2 sign(A^{\pm}) \frac{df^{\pm}}{dn^{\pm}} dn^{+}d\widetilde{R}^{\pm} .
\end{equation}
\par From the continuity of the metric and $ \widetilde{R} $ on $ \Sigma_0 $ we get that $\frac{df^{+}}{dn^{+}}(0)=\pm \frac{df^{-}}{dn^{-}}(0)$ when $C^{+}_{0}=C^{-}_{0}$ . It means that the two given vacua can only differ in conformal factors $r^{\pm}(n^{\pm},\widetilde{R}^{\pm})$, but they must also coincide up to the second order in $n^{\pm}$ on $\Sigma_0$:
\begin{equation} \label{84}
    r^{+}(0,\widetilde{R}^{+})=r^{-}(0,\widetilde{R}^{-}), \quad \partial_{n^{+}}r^{+}(0,\widetilde{R}^{+})=\partial_{n^{-}}r^{-}(0,\widetilde{R}^{-}).
\end{equation} 

\subsection{Vaidya-type solutions}
\par Let's proceed to the study of null thin shells separating the Vaidya-type metric as $\Omega^{+}$ and the vacuum or another Vaidya-type metric as $\Omega^{-}$. These Vaidya-type solutions for spherically symmetric conformal gravity were also obtained in the paper Ref.~\refcite{Berezin2016}.

\par  The two-dimensional “metric” of the outgoing  Vaidya-type solution chosen as $\Omega^{+}$ in the coordinates $\{\widetilde{R}^{+},u^{+}\}$ is: \begin{equation} \label{85}
    d\widetilde{s}^{+ 2}_{2}=A^{+}(\widetilde{R}^{+},u^{+}) du^{+2}+2du^{+}d\widetilde{R}^{+}, 
\end{equation} 
where $A^{+}(\widetilde{R}^{+},u^{+})=\frac{1}{6}(\widetilde{R}^{+ 3}-12\widetilde{R}^{+}+C^{+}_{0}(u^{+}))$, $C^{+}_{0}(u^{+})$ is an arbitrary function of variable $u^{+}$.
\par Two types of spherically symmetric lightlike hypersurfaces exist for this spacetime. A hypersurface defined by an arbitrary function of $u^{+}$ is the first type. The second type is a hypersurface given by any function $F^{+}(u^{+},\widetilde{R}^{+})$ that obeys the following equation:
\begin{equation} \label{86}
   \partial_{u^{+}}F^{+}= \frac{1}{12}(\widetilde{R}^{+ 3}-12\widetilde{R}^{+}+C^{+}_{0}(u^{+}))    \partial_{\widetilde{R}^{+}}F^{+} .
\end{equation}
\par In what follows, the second class of surfaces will be used because in theory it could represent a thin shell with radiation which is transversal to $\Sigma_0$.
\par Let's choose a vacuum with constant $\widetilde{R}^{-}=\pm 2$ as $\Omega^{-}$. From the continuity of the two-dimensional scalar curvature it follows that $n^{+}(x^{+})=\widetilde{R}^{+}\pm 2$ but according to (\ref {86}) a hypersurface of this type can be lightlike only if $C^{+}_{0}(u^{+})=\pm 16$. In this case a Vaidya-type solution boils down to a vacuum with variable $\widetilde{R}$ and it means that such matching does not exist.

\par If the $\Omega^{-}$ is a vacuum with variable $\widetilde{R}$ we can assume that the hypersurface equation is given by an arbitrary function of variable $u^{-}$: $n^{-}(x^{-})=F^{-}(u^{-})=0$. Let's use the coordinates $\{n^{\pm},\widetilde{R}^{\pm}\}$ in the $\Omega^{\pm}$ when the two-dimensional “metrics” of the corresponding domains have the following form:
\begin{equation} \label{87}
    d\widetilde{s}_{2}^{- 2}=A^{-} \left(\frac{df^{-}}{dn^{-}} \right)^{2} dn^{- 2}+2 sign(A^{-}) \frac{df^{-}}{dn^{-}} dn^{-}d\widetilde{R}^{-},
\end{equation}
\begin{equation} \label{88}
    d\widetilde{s}_{2}^{+ 2}=A^{+} \left(\partial_{u+} F^{+} \right)^{-2} dn^{+ 2}-2 \left(\partial_{u+} F^{+} \right)^{-1} dn^{+}d\widetilde{R}^{+}.
\end{equation}
Here $f^{-}(n^{-})$ denote a function inverse to $F^{-}(u^{-})$ and $F^{+}(u^{+},\widetilde{R}^{+})$ is a function defined by (\ref{86}).
\par From the continuity of the metric and two-dimensional scalar curvature on the hypersurface we get that $\frac{df^{-}}{dn^{-}}(0)=\pm \partial_{u+} F^{+}(\widetilde{R}^{+},u^{+})|_{\Sigma_0}$ and $A^{+}(\widetilde{R}^{+},u^{+})|_{\Sigma_0}=A^{-}(\widetilde{R}^{-})|_{\Sigma_0}$. Whence it follows that $C^{+}_{0}(u^{+})|_{\Sigma_0}=C^{-}_{0}|_{\Sigma_0}=const$. In this case $\Sigma_0$ in $ \Omega^{+} $ belongs to the first type of null hypersurfaces in the Vaidya-type metric and this type corresponds to the absence of radiation.
\par Finally, let's consider null thin shells separating two Vaidya-type metrics. The hypersurface equations in the domains $\Omega^{\pm}$ in this case are functions $n^{\pm}(x^{\pm})=F^{\pm}(u^{\pm},\widetilde{R}^{\pm})$ given by the equation (\ref{86}). Using the coordinates $\{n^{\pm},\widetilde{R}^{\pm}\}$ we get:
 \begin{equation} \label{90a}
    ds_{2}^{\pm 2}=A^{\pm} \left(\partial_{u\pm} F^{\pm} \right)^{-2} dn^{\pm 2}-2 \left(\partial_{u\pm} F^{\pm} \right)^{-1} dn^{\pm}d\widetilde{R}^{\pm} .
\end{equation}
\par The following relations are a consequence of the continuity of the metric and the two-dimensional scalar curvature on the hypersurface:
$$\partial_{u-} F^{-}(\widetilde{R}^{-},u^{-})|_{\Sigma_0}= \partial_{u+} F^{+}(\widetilde{R}^{+},u^{+})|_{\Sigma_0}, \quad
A^{+}(\widetilde{R}^{+},u^{+})|_{\Sigma_0}=A^{-}(\widetilde{R}^{-},u^{-})|_{\Sigma_0},
$$
which in turn implies that : $C^{+}_{0}(u^{+})|_{\Sigma_0}=C^{-}_{0}(u^{-})|_{\Sigma_0}$. Thus the two spacetimes under consideration can differ only by the conformal factors $r^{\pm}(n^{\pm},\widetilde{R}^{\pm})$ starting from the second derivative with respect to $n^{\pm}$, since the first derivatives must be continuous due to the Lichnerowicz conditions:
\begin{equation} \label{90b}
    r^{+}(0,\widetilde{R}^{+})=r^{-}(0,\widetilde{R}^{-}), \quad \partial_{n^{+}}r^{+}(0,\widetilde{R}^{+})=\partial_{n^{-}}r^{-}(0,\widetilde{R}^{-})\, .
\end{equation}
\par The hypersurface could be of the second type in this situation, but it still cannot be interpreted as a thin shell with radiation.

\section{Weakening of the Lichnerowicz conditions in spherical symmetry}

\par Let us study the existence of models in quadratic gravity for which the Lichnerowicz conditions can be weakened or eliminated in the framework of spherically symmetric metrics and hypersurfaces.
\par It was previously shown that the quadratic gravity Lagrangian can be expressed as a combination of the square of the scalar curvature, the square of the Weyl tensor and the Gauss-Bonnet term. Using the relations presented in the papers Ref.~\refcite{Berezin2016}, Ref.~\refcite{Shapiro} for a spherically symmetric metric  (\ref{52}) we obtain the following:
\begin{equation} \label{spher}
\sqrt{|g|} \, C^2=\frac{1}{3}\,\sin{\theta} \sqrt{|\widetilde{\gamma }|}\left ( \widetilde{R}-2 \right )^{2}\, ,\quad \sqrt{|g|} \, R^2=\sin{\theta} \sqrt{|\widetilde{\gamma }|}\left ( \widetilde{R}-2 -\frac{6}{r}\, \widetilde{\sigma }\right )^{2} \, ,    
\end{equation}
\begin{multline}
\sqrt{|g|}\, GB=\sin \theta \, \sqrt{|\widetilde{\gamma }|}\, \left\{ \right.\frac{4}{r^{2}}\,  \widetilde{R}\, \left ( 3\widetilde{\Delta } -r^{2}-2r\, \widetilde{\sigma }\right )+\frac{8}{r^{3}}\, \widetilde{\sigma }\left ( r\, \widetilde{\sigma }-\widetilde{\Delta } \right )-\frac{24}{r^{4}}\, \widetilde{\Delta }^{2}+\\+\frac{32}{r^{3}}\, \partial _{\alpha }r\, \partial _{\beta }r\, \widetilde{\triangledown }^{\alpha }\widetilde{\triangledown }^{\beta }r-\frac{8}{r^{2}}\,\widetilde{\triangledown }_{\alpha }\widetilde{\triangledown }_{\beta }r\:  \widetilde{\triangledown }^{\alpha }\widetilde{\triangledown }^{\beta }r \left. \right \} \, .
\end{multline}
\par In the fourth section, it was noted that the Gauss-Bonnet term does not contribute to the motion equations not only in the bulk but also on the boundary. Therefore, it can be excluded from the original quadratic gravity Lagrangian. Strictly speaking, this point requires more detailed study because the Lichnerowicz conditions were used in the derivation of (\ref{30a}-\ref{30c}) and here we are exploring the situations where they can be weakened or removed completely. In this regard, let us clarify that we consider problems where singular hypersurface arises as the limit of some non-singular matter distribution in the $\Omega$ spacetime without singularities on the boundary $\partial \Omega$ for which the above assumption can be applied.
\par Substituting the expressions (\ref{spher}) into the original action for quadratic gravity without the Gauss-Bonnet term and integrating over the angles, we obtain a two-dimensional effective action:
\begin{multline}
S_{2q}=-\frac{1}{4}\int_{\Omega } \sqrt{|\widetilde{\gamma }|}\left\{ \right.\frac{1}{2}\left ( 2\alpha_3+2\alpha_1+\alpha_2 \right )\left ( \widetilde{R}-2 \right )^{2}+\left ( \widetilde{R} -2\right )\left ( \alpha _{4}\,  r^{2}-12\, \beta\,  \frac{\widetilde{\sigma }}{r}\, \right )+\\+36\, \beta \, \frac{\widetilde{\sigma }^{2}}{r^{2}}-6\, \alpha _{4}\, r\, \widetilde{\sigma }+\alpha_5\,r^{4} \Lambda \left. \right \}d^{2}x=-\frac{1}{4 }\int_{\Omega }\sqrt{|\widetilde{\gamma }|}L_{2q}\, d^{2}x\, ,
\end{multline}
where $ \beta = \alpha _{3}+ \frac{1}{3}\alpha_{1}+\frac{1}{3}\alpha_{2}$ .
\par If we set $\alpha_3=-\alpha_1-\frac{1}{2}\alpha_2$, then the resulting action will be linear in $\widetilde{R}$. In this case the existence of a jump in the derivatives of $\widetilde{\gamma}_{\alpha \beta}$ and a delta function in the curvature could be possible but the product of the delta function and the theta function forbidden in the conventional theory of distributions still appears.
\par On the other hand, it follows from the definition of the invariants $\widetilde{R}$ and $\widetilde{\sigma}$ that they contain only the first derivatives with respect to $n$ for lightlike hypersurfaces, since $\gamma^{ nn}|_{\Sigma_0}=N_{a}N_{b}g^{ab}=0$ for this case. Thus, for spherically symmetric null hypersurfaces the presence of jumps in the derivatives of the metric does not lead to the appearance of a delta function in $\widetilde{R}$, $\widetilde{\sigma}$ and in the Lagrangian $L_{2q}$.
\par This reasoning does not work for timelike (spacelike) hypersurfaces, because for them the spherical geometry invariants must include the second derivatives with respect to $n$:
$$
[\widetilde{\sigma}]=\varepsilon \,r^2 [\partial^{2}_{nn}r],\quad [\widetilde{R}]=2\varepsilon\,r^2[\partial_n K]+6\varepsilon r [\partial^{2}_{nn}r] \, ,
$$
where $K=-\triangledown_a\,N^a$ is the trace of the hypersurface extrinsic curvature tensor.
\par Let's consider a null singular hypersurface given by the equation $n(x)=0$ requiring only the continuity of the metric on $\Sigma_0$. In this case, the above invariants can be represented in the following form:
$$
\widetilde{R}=\widetilde{R}^{+}\, \theta (n(x))+\widetilde{R}^{-}\, \theta (-n(x)),\quad
\widetilde{\sigma }=\widetilde{\sigma }^{+}\, \theta (n(x))+\widetilde{\sigma }^{-}\, \theta (-n(x)) .
$$
Substituting these expressions into the two-dimensional effective action, we get:
$$
S_{2q}=-\frac{1}{4 }\int_{\Omega_{2}^{+} }\sqrt{|\widetilde{\gamma }|}L^{+}_{2q}\, d^{2}x-\frac{1}{4 }\int_{\Omega_{2}^{-} }\sqrt{|\widetilde{\gamma }|}L^{-}_{2q}\, d^{2}x\, .
$$
\par Let's vary the resulting action with respect to the two-dimensional “metric” $\widetilde{\gamma }_{\alpha \beta }$ and the radius $r$, which acts as a dilaton here. According to the properties of the variational derivative:
\begin{multline}
\delta S_{2q}=-\frac{1}{4 }\int_{\Omega_{2}^{+} }\sqrt{|\widetilde{\gamma }|}\left \{ \frac{\delta L^{+}_{2q}}{\delta \widetilde{\gamma }^{\alpha \beta }}\, \delta \widetilde{\gamma }^{\alpha \beta }-\frac{1}{2} \widetilde{\gamma }_{\alpha \beta }\, L^{+}_{2q}\, \delta \widetilde{\gamma }^{\alpha \beta }+\frac{\delta L^{+}_{2q}}{\delta r}\, \delta r\right \}\, d^{2}x\,-\\ -\frac{1}{4 }\int_{\Omega_{2}^{-} }\sqrt{|\widetilde{\gamma }|}\left \{ \frac{\delta L^{-}_{2q}}{\delta \widetilde{\gamma }^{\alpha \beta }}\, \delta \widetilde{\gamma }^{\alpha \beta }-\frac{1}{2} \widetilde{\gamma }_{\alpha \beta }\, L^{-}_{2q}\, \delta \widetilde{\gamma }^{\alpha \beta }+\frac{\delta L^{-}_{2q}}{\delta r}\, \delta r\right \}\, d^{2}x\, .
\end{multline}
\par In the action variation we single out the terms that appear on the hypersurface after the application of the Stokes' theorem:
$$
\delta S_{2q}|_{\Sigma_0}=-\frac{1}{4}\int_{\Omega_{2}^{+} }\sqrt{|\widetilde{\gamma }|}\, \widetilde{\triangledown} _{\alpha } V_{2 }^{+\alpha }\, d^{2}x\,-\frac{1}{4 }\int_{\Omega_{2}^{-} }\sqrt{|\widetilde{\gamma }|}\, \widetilde{\triangledown}_{\alpha } V_{2 }^{-\alpha }\, d^{2}x\,,
$$
\begin{multline}
V_{2}^{+\alpha }=\left ( \widetilde{\gamma }^{\beta \nu }\,  \widetilde{\gamma }^{\alpha \mu }-\widetilde{\gamma }^{\mu \nu }\, \widetilde{\gamma }^{\alpha \beta }\right )\left \{ X^{+2} \widetilde{\triangledown }_{\beta }\left ( \frac{\delta \widetilde{\gamma }_{\mu \nu }}{X^{+}} \right )-3\partial _{\beta }r\, Y^{+}\, \delta \widetilde{\gamma }_{\mu \nu }\right \}-\\-3\, \widetilde{\partial }^{\nu }r\, \widetilde{\gamma }^{\alpha \mu }Y^{+}\, \delta \widetilde{\gamma }_{\mu \nu }+6Y^{+2}\, \widetilde{\triangledown }^{\alpha }\left ( \frac{\delta r}{Y^{+}} \right ),\\ X^{+}=\frac{\partial L_{2q}^{+}}{\partial \widetilde{R}^{+}}=\alpha _{4}\, r^{2}-\frac{12\beta }{r}\, \widetilde{\sigma }^{+}+\left ( 2\alpha_3+2\alpha_1+\alpha_2 \right )\left ( \widetilde{R}^{+} -2\right ), \\ Y^{+}=\frac{1}{6}\frac{\partial L_{2q}^{+}}{\partial \widetilde{\sigma }^{+}}=-\frac{2\beta }{r}\left ( \widetilde{R}^{+} -2\right )+\frac{12\beta }{r^{2}}\widetilde{\sigma }^{+}-\alpha_4 \, r\,=-2\beta\,  r\, R^{+}-\alpha_4 \, r\, .
\end{multline}
The vector $V_{2}^{-\alpha }$ as well as the scalars $X^{-}$ and $Y^{-}$ are defined similarly to the $V_{2}^{+\alpha }$, $X^{+}$, $Y^{+}$ .
\par Let's move on to the coordinates $\{n,\widetilde{\lambda}\}$, which are related to the original ones by changing the variable $\lambda$: $\widetilde{\lambda} =\int \frac{d\lambda }{r^{ 2}(0,\lambda )} $. As a result of this replacement, we get: $\widetilde{\gamma}_{n \widetilde{\lambda}}(0,\lambda)=1$. 
\par The components of the two-dimensional “metric” $\widetilde{\gamma}_{ij}$ are continuous on $\Sigma_0$ in these coordinates. In addition, it is necessary to require the continuity of the radius on the hypersurface: $r^{-}(0,\widetilde{\lambda })=r^{+}(0,\widetilde{\lambda })$ .
\par Only the $n$-th component of the vector $[V_{2}^{\alpha }]$ is present in the motion equations of the $\Sigma_0$. Discarding the terms that are total divergence on the hypersurface, we get:
\begin{multline}
[V_{2}^{n }]= -\left (\, 6\, [\partial _{n}r\, Y]+3\, \widetilde{\gamma}^{\widetilde{\lambda} \widetilde{\lambda} }\, \partial _{\widetilde{\lambda} }r\, [Y] +[\partial _{n}X] +\partial _{\widetilde{\lambda} }\widetilde{\gamma }_{nn}\, [X]\, \right )\delta \widetilde{\gamma }_{\widetilde{\lambda} \widetilde{\lambda} }+\\+[X\, \partial _{n}\, \delta \widetilde{\gamma }_{\widetilde{\lambda} \widetilde{\lambda} }]+2[\partial _{\widetilde{\lambda} }X]\, \delta \widetilde{\gamma }_{n \widetilde{\lambda} }-12[\partial _{\widetilde{\lambda} }Y]\, \delta r.
\end{multline}
It should be clarified that in this model, the jumps of a certain variable on the hypersurface $\Sigma_0$ are determined taking into account the first derivatives with respect to $n$ as well.
\par In a way similar to the gravitational action, we integrate over the angles and single out the surface part in the variation of the matter action:
\begin{multline}
\delta S_{2m}|_{\Sigma_0}=-2\pi \int _{\Sigma _{0}}r^{4}\sqrt{\widetilde{\gamma }(0,\widetilde{\lambda} )}\, \left ( S^{\alpha \beta } \delta \gamma _{\alpha \beta }+2\, S^{22}\, \delta g_{22}\right )d\widetilde{\lambda}=\\ =-2\pi \int _{\Sigma _{0}}\sqrt{\widetilde{\gamma }(0,\widetilde{\lambda} )}\, \left ( r^{6}\, S^{\alpha \beta } \delta \widetilde{\gamma } _{\alpha \beta }+2 r^{3} S\,  \delta r\right )d\widetilde{\lambda}\, ,
\end{multline}
where $S=r^{2}\left ( S^{\alpha \beta }\, \widetilde{\gamma }_{\alpha \beta }-2\, S^{22} \right )$ is a trace of the surface energy-momentum tensor.
\par The surface part of the motion equations is obtained from the principle of least action:
\begin{multline} \label{meq}
-\left (\, 6\, [\partial _{n}r\, Y]+3\, \widetilde{\gamma}^{\widetilde{\lambda} \widetilde{\lambda} }\, \partial _{\widetilde{\lambda} }r\, [Y] +[\partial _{n}X] +\partial _{\widetilde{\lambda} }\widetilde{\gamma }_{nn}\, [X]\, \right )\delta \widetilde{\gamma }_{\widetilde{\lambda} \widetilde{\lambda} }+\\+[X\, \partial _{n}\, \delta \widetilde{\gamma }_{\widetilde{\lambda} \widetilde{\lambda} }]+2[\partial _{\widetilde{\lambda} }X]\, \delta \widetilde{\gamma }_{n \widetilde{\lambda} }-12[\partial _{\widetilde{\lambda} }Y]\, \delta r=8\pi \left ( r^{6}\, S^{\alpha \beta } \delta \widetilde{\gamma } _{\alpha \beta }+2 r^{3} S\,  \delta r\right ) .
\end{multline}
\par If we require the continuity of the scalars $X$ and $Y$ on $\Sigma_0$ in the equation  (\ref{meq}), then we get  (\ref{54}) in the variables $\{n,\widetilde{\lambda}\}$. Moreover, from  (\ref{meq}) it follows that even for such a model of null singular hypersurface $S^{nn}=0$.
\par Here the term $[\partial _{n}\, \delta \widetilde{\gamma }_{\lambda \lambda }]\neq 0$ can no longer be bracketed, as it was done before. However, due to the implicit presence of the derivative of the delta function in the motion equations, it is still expressed through the combination of $\delta g_{ij}$ with arbitrary functions as coefficients:
\begin{equation}
[X\, \partial _{n}\, \delta \widetilde{\gamma }_{\widetilde{\lambda} \widetilde{\lambda} }]=B_{\widetilde{\lambda} \widetilde{\lambda}}^{\widetilde{\lambda} \widetilde{\lambda}}(\widetilde{\lambda},r)\, \delta \widetilde{\gamma }_{\widetilde{\lambda} \widetilde{\lambda}}+B_{\widetilde{\lambda} \widetilde{\lambda}}^{2 2}(\widetilde{\lambda},r)\, \delta r \, .
\end{equation}
\par Let's also investigate thecase of conformal gravity, for which: $\beta=\alpha_4=0, \quad \alpha_3=\frac{1}{3}\alpha_1, \quad \alpha_2=-2 \alpha_1 $ in order to compare this model with the previous results. \par From (\ref{meq}) with the appropriate coefficients, we obtain the motion equations for spherically symmetric null double layer in conformal gravity:
\begin{equation} \label{eq1}
[\partial _{\widetilde{\lambda}} \widetilde{R}]=\frac{12\pi }{\alpha _{1}} \, r^{6}\, S^{n \widetilde{\lambda}},
\end{equation}
\begin{equation}  \label{eq2}
[\partial _{n}\widetilde{R}] +\partial _{\widetilde{\lambda} }\widetilde{\gamma }_{nn}\, [\widetilde{R}]-B_{\widetilde{\lambda} \widetilde{\lambda}}^{\widetilde{\lambda} \widetilde{\lambda}}(\widetilde{\lambda},r)=-\frac{12\pi }{\alpha _{1}} \, r^{6}\, S^{\widetilde{\lambda} \widetilde{\lambda} },
\end{equation}
\begin{equation} \label{trace}
B_{\widetilde{\lambda} \widetilde{\lambda}}^{2 2}(\widetilde{\lambda},r)=16\pi   r^{3} S =0\, .
\end{equation}
\par It is known that the trace of the energy-momentum tensor is equal to zero for conformal gravity. \cite{Berezin2016} It means that the trace of the bulk parts of the energy-momentum tensor is zero: $T^{\pm}_{ab} \, g^{ab}=0$. However the trace of the surface tensor energy-momentum $S^{a}_{a}$ is not necessary zero, since we did not assume by default that conformal invariance is preserved directly on $\Sigma_0$. Nevertheless, if our model is a limiting case of some non-singular distribution of matter, then it makes sense to demand explicitly: $S^{a}_{a}=0$.
\par Only the first equation from the system  (\ref{eq1}-\ref{trace}) determines the dynamics of the null double layer, since the other two define unknown functions: $B_{\widetilde{\lambda} \widetilde{\lambda}}^{2 2}(\widetilde{\lambda},r),\; B_{\widetilde{\lambda} \widetilde{\lambda}}^{\widetilde{\lambda} \widetilde{\lambda}}(\widetilde{\lambda},r).$ 
\par It follows from the equation  (\ref{trace}) that $S^{22}=S^{n \widetilde{\lambda}}$. This means in particular that the null double layer cannot consist of dust, for which $S^{2}_{2}=S^{3}_{3}=0$ in the spherically symmetric case. It makes sense since dust particles cannot move at the speed of light. 
\par On the other hand, a null double layer with $S^{\widetilde{\lambda} \widetilde{\lambda}}=0$ that simulates vacuum burning \cite{Berezin83} is possible. In this case, there is a connection between the so-called entropy source which is part of $S^{22}$ and the radiation from the double layer, presumably given in terms of $S^{n \widetilde{\lambda}}$. The concept of an entropy source was first introduced in the above article Ref.~\refcite{Berezin83}.
\par The double layer exists only if the jump $[\partial _{\widetilde{\lambda}} \widetilde{R}]$ is non-zero. This automatically implies that the matching of two vacua with constant $\widetilde{R}$ does not create a double layer.
\par For a vacuum with a variable $\widetilde{R}$ described by the two-dimensional “metric”  (\ref{79}) the null hypersurface is an arbitrary function of the variable $u$ or $v$. Let's choose a hypersurface defined by an arbitrary function of $u$: $n=F(u)$, then $\widetilde{\lambda}=\int \frac{df}{dn}(0)\, \widetilde{H}( f(0),v)\, dv$. In the coordinates $\{n,\widetilde{\lambda}\}$ the two-dimensional “metric” has the form:
$$
ds_{2}^{+2}=2\, \frac{\widetilde{H}(n,\widetilde{\lambda })\, \frac{df}{dn}(n)}{\widetilde{H}(0,\widetilde{\lambda })\, \frac{df}{dn}(0)}dn\, d\widetilde{\lambda },
$$
where $f(n)$ is a function inverse to $F(u)$.
\par Thus for a vacuum with the variable $\widetilde{R}$ the following holds:
\begin{equation} \label{curvder}
 \partial _{\widetilde{\lambda }}\widetilde{R} |_{\Sigma _{0}}=\left(\partial _{v}\widetilde{R}\, \frac{dv}{d\widetilde{\lambda }}\right) |_{\Sigma _{0}}=\frac{1}{\frac{df}{dn}(0)} \, .
\end{equation}
\par For the matching of two vacua, namely, the vacuum with the variable $\widetilde{R}$ and the vacuum with the constant $\widetilde{R}$ or two vacua with the variable $\widetilde{R}$, the dynamics of the double layer is given by the equation:
\begin{equation}
\frac{12\pi }{\alpha _{1}} \, r^{6}\, S^{n \widetilde{\lambda}  }=\frac{12\pi }{\alpha _{1}} \, r^{4}\, S^{n \lambda}=const.
\end{equation}
\par As an example, let's consider a null double layer separating the Schwarzschild spacetime as $\Omega^{+}$ and the Minkowski vacuum as $\Omega^{-}$. Two-dimensional “metric” in $\Omega^{+}$ is:
$$
ds_{2}^{+2}=\frac{1}{r^{2}(u,v)}\left ( 1-\frac{r_{g}}{r(u,v)} \right )du\, dv \, .
$$
\par Let's choose a null hypersurface $u=u_0$ for which: $n=u-u_0,\quad \widetilde{\lambda }=\int \frac{dv}{2\, r^{2}(u_0,v )}\left ( 1-\frac{r_g}{r(u_0,v)} \right )$ then:
\begin{equation} \label{Schwarz}
\partial_{\widetilde{\lambda }} \widetilde{R}\, |_{\Sigma_{0} }=\partial_{\widetilde{\lambda }}\left ( 2-\frac{6\, r_g}{r} \right )\, |_{\Sigma_{0} }=\frac{6 r_g}{r^2}\, \partial _{v}r\, \frac{dv}{d\widetilde{\lambda }}\, |_{\Sigma_{0} }=6\, r_g .
\end{equation}
\par For this case the dynamics of a double layer is defined by:
\begin{equation}
r^{4}(\lambda )=\frac{r_g}{2\pi }\alpha _{1}\left ( S^{n\lambda } (\lambda )\right )^{-1}.
\end{equation}
\par Let's also clarify that (\ref{Schwarz}) differs from  (\ref{curvder}) because the coordinates $\{u,v\}$ in the formula (\ref{79}) for a vacuum with variable $\widetilde {R}$ do not match the standard double null coordinates for the Schwarzschild metric:
$$
ds_{2}^{+2}=\frac{1}{r^{2}(u,v)}\left ( 1-\frac{r_{g}}{r(u,v)} \right )du\, dv=\frac{du\, dv}{\left (6 r_g  \right )^{2}}\, \frac{1}{6}\left ( \widetilde{R}^{3} -12\widetilde{R}+16\right ) \,,
$$
namely, a replacement $\frac{u}{6 r_g}\mapsto u,\quad \frac{v}{6 r_g}\mapsto v\, $ is necessary to bring the two-dimensional “metric” for the Schwarzschild spacetime to the form (\ref{79}).
\par Taking into account the uncertainty in the definition of the normal to a null hypersurface, it makes sense to rewrite the motion equation (\ref{eq1}) in the invariant form:
\begin{equation}
    N_{a}\, \partial ^{a} \widetilde{R}=\frac{6\pi }{\alpha _{1}}\, r^{2}\, S^{\alpha \beta }\, \gamma _{\alpha \beta }=\frac{6\pi }{\alpha _{1}}\, r^{2}\, S^{b c } N_{b} \, l_{c} \, .
\end{equation}
\par In contrast to spherically symmetrical null thin shells in conformal gravity, there is a null double layer corresponding to the matching between Vaidya-type metric and a vacuum with constant $\widetilde{R}$. The motion equation for this case is:
\begin{equation}
    \pm \partial_{u}F(x)=\frac{6\pi }{\alpha _{1}}\, r^{4}\, S^{\alpha \beta }(x)\, \gamma _{\alpha \beta }(x) \, ,
\end{equation}
where $\{x^{\alpha},\theta, \phi\}$ are arbitrary coordinates that are continuous in the neighborhood of $\Sigma_0$, $F(x)$ is a function given by the equation  (\ref{86}). The upper sign corresponds to the case when $\Omega^{-}$ is a Vaidya-type metric, $\Omega^{+}$ is the vacuum with constant $\widetilde{R}$, and the lower sign corresponds to the reverse situation.

\section{Conclusion}

\par It was demonstrated that not only for timelike and spacelike singular hypersurfaces \cite{Berezin2020} but also for null singular hypersurfaces both in general relativity and in quadratic gravity the field equations can be derived using exclusively the least action principle.

\par The coefficient combinations $\beta_1,\, \beta_2$ present in the motion equations of singular hypersurface of the most general type in quadratic gravity are zero for the quadratic Gauss-Bonnet term. It means that for this particular case there are no double layers or thin shells if the Lichnerowicz conditions are met.

\par Unlike timelike and spacelike double layers null double layer in quadratic gravity does not have the so-called “external pressure”. In addition, we can suppose that the lightlike double layer radiates, since for this type of null singular hypersurfaces the “external flow” which is three-dimensional vector $S^{ni}=S^{i}_{a}\, N^{a}$ presumably associated with radiation cannot be zero.

\par As it turns out, the criterion for a hypersurface to be a thin shell rather than a double layer, which is the most general case of a singular hypersurface in quadratic gravity, depends on the type of hypersurface. For a null thin shell, it is the absence of a jump in scalar curvature; for the rest, it is the absence of jumps in the components of the Ricci tensor.

\par It was demonstrated that for null singular spherically symmetric hypersurfaces in quadratic gravity, the Lichnerowicz conditions imply a continuity of the scalar curvature $[R]=0$. This means that such singular hypersurface can only be a thin shell. In this case the system of field equations (\ref {30e}) reduces to one that is expressed in terms of spherical geometry invariants along with the Lichnerowicz conditions.

\par On the other hand, if we consider the effective action of quadratic gravity without the Gauss-Bonnet term for the case of spherical symmetry, then it turns out that the Lichnerowicz conditions can be removed completely for null singular hypersurfaces. In this model a spherically symmetric null double layer exist end it doesn't have “external pressure” as well.

\par Spherically symmetric null thin shells and double layers separating two spherically symmetric solutions of conformal gravity were explored as an application. Namely, the matchings between spherically symmetric vacuum and Vaidya-type solutions were investigated.

\par For thin shells the continuity of the two-dimensional scalar curvature imposes certain restrictions on the possibility of matching. As a result there are only four types of matchings for the metrics considered here. The first type is a matching between a vacuum with a constant $\widetilde{R}$ and a vacuum with a variable $\widetilde{R}$ where $\Sigma_0$ turned out to be an analogue of the double horizon in the metric with variable $\widetilde{R}$. The second type is a matching of two vacuums with a variable $\widetilde{R}$. It exists only if the metrics coincide up to a conformal factor. The third type is a matching between Vaidya-type solution and a vacuum with variable $\widetilde{R}$ where the resulting null thin shell is in fact a singular part of the Vaidya-type spacetime. The fourth type is matching between two Vaidya-type metrics and it also exists only if they coincide up to a conformal factor. The lack of radiation in the last two cases is presumably due to the absence of “external flow” for thin shells.
\par For a spherically symmetric null double layer in the conformal gravity only the matching of two vacua with the constant $\widetilde{R}$ does not exist. All other options for matching vacuum solutions and Vaidya-type solutions are available.

\end{document}